\documentclass[final,5p,times,twocolumn]{elsarticle}

\usepackage{amssymb}
\usepackage{amsmath}
\usepackage{mathtools}
\usepackage{subfigure}
\usepackage[ruled,vlined]{algorithm2e}
\usepackage{graphicx}
\usepackage{caption}
\usepackage{float}

 \usepackage{multicol}
 \usepackage[switch]{lineno}

\begin{document}

\begin{frontmatter}



\title{Origin and control of pseudo-rotating spiral jets} 


\author[label1]{Karol Wawrzak}
\ead{karol.wawrzak@pcz.pl}
\author[label2,label3]{Yiqing Li}
\author[label2]{Bernd R. Noack}
\author[label1]{Artur Tyliszczak}

\address[label1]{Faculty of Mechanical Engineering, Czestochowa University of Technology, 42-201 Czestochowa, Poland}
\address[label2]{Chair of Artificial Intelligence and Aerodynamics, School of Mechanical Engineering and Automation, \\Harbin Institute of Technology, 518055 Shenzhen, PR China}
\address[label3]{Department of Mechanical Engineering, University College London, London, UK}

\date{}

\journal{Elsevier}

\begin{abstract}
Active flow control (AFC) methods for jet-type flows have been extensively explored since the 1970s. Spectacular examples demonstrating the AFC power and the beauty of fluid mechanics include bifurcating and blooming jets. Recent advances in machine learning-based optimization have enabled efficient exploration of high-dimensional AFC, revealing control solutions beyond human intuition. 
The present paper focuses on one such discovery: the pseudo-rotating spiral jet. This phenomenon manifests as separate branches disconnected from the main jet stream, formed by vortical structures aligned along curved paths rotating around the initial jet axis. We investigate the origin of these jet-type patterns and formulate new rules for their control, showing that spiral jets belong to a family of multi-armed jets observable only at specific control settings. 
Furthermore, we demonstrate how human perception of three-dimensional imagery depends on the observable domain and vortex lifetime.
Notably, the apparent rotation of spiral arms—despite having a well-defined frequency—is an illusion arising from the tendency to connect neighboring moving objects into continuous patterns.  In contrast to  the chaotic behavior of small-scale turbulence, we show that large-scale flow motion resulting from AFC operating in a deterministic manner is only seemingly unpredictable. Through theoretical analysis and 3D simulations, we develop a remarkably simple yet precise kinematic model that captures the formation and motion of these vortical paths. This model replicates the outcomes of complex flow simulations, reproduces the apparent jet shape, and facilitates the identification of the actual pattern. The findings offer new perspectives for both academic researchers and industrial engineers.
\end{abstract}




\end{frontmatter}


\section{Introduction}
\label{sec:headings}


Fluid mechanics is rich in important and fascinating phenomena. Among them, jets represent a canonical flow type that, under certain conditions, exhibits astonishing dynamics. Researchers have been studying jets for decades, driven both by scientific curiosity and their widespread occurrence in various interdisciplinary applications.
Jets play a crucial role in numerous technical and everyday contexts. In engineering, they are essential for fuel delivery in combustion chambers and for heat and air distribution in heat exchangers and air conditioning systems. In medicine, jet-based devices such as nasal sprays and injectors aid in treatment and recovery. In daily life, they enhance convenience and comfort, appearing in perfume atomizers, fire sprinklers, and garden irrigation systems.
Many more examples of jet applications could be listed, highlighting the importance of understanding their dynamics. As a result, jet flows have been the subject of extensive theoretical, experimental, and numerical research. Historical overviews and comprehensive discussions on this topic can be found in review papers \citep{Lipari_2011,Ball_Pollard_2012,Kaushik_2015,Boguslawski_et_al_IJNMHFF_2016}.

Regardless of the application, jets are streams of fluid that share common characteristics upon issuing from a nozzle. Near its vicinity, they are dominated by large-scale coherent structures appearing as toroidal vortices (vortical puffs). These structures originate in the shear layer due to the Kelvin-Helmholtz instability and emerge at a specific frequency known as the preferred mode \citep{GutmarkHo_PoF_1983,Ball_Pollard_2012,Boguslawski_et_al_JFM_2019}.
The strength and shape of these vortices depend on small disturbances that inevitably exist at nozzle edges due to imperfections in the manufacturing process, such as asymmetry and surface roughness, or instability of a power supply system (a pump). As the vortices travel downstream, they rotate and interact with neighbouring ones, potentially generating small-scale side jets, connecting through long rib-like streamwise vortices, or undergoing vortex pairing. Eventually, at a certain distance from the nozzle, these vortices break up, leading to a fully developed turbulent flow characterized by a broad range of small-scale vortices.
All these phenomena are strongly influenced by the origin and magnitude of disturbances. If they affect the jet in an unorganized, stochastic manner (randomness of the incoming stream), their impact on jet dynamics remains limited, resulting in a general similarity among jets downstream. However, if these disturbances are assumed to achieve specific objectives and controlled in real-time based on current flow conditions, their coordinated actions can lead to remarkable outcomes. 

The above description introduces the science of flow control, which plays a crucial role in jet applications.  
Despite significant knowledge of the jets dynamics, controlling them remains an active area of research, employing both passive and active techniques~\citep{Gad-el-Hak_book_2000,Collis_ProgAeroSci_2004,Ashill_Fulker_Hackett_2005,Cattafesta_AnnRev_2011,Ali_2021}. Depending on the application, the goal of control varies and may rely on vectoring the jet momentum, changing the jet shape or size, or improving mixing between the jet and the surrounding medium. 

The passive flow control (PFC) approach involves adding fixed elements to the nozzle (vortex generators, orifice plates) or modifying its shape. Most of the research in this direction concentrates on non-circular jets emanating from elliptical or polygonal nozzles (e.g. square, rectangular, triangular)~\citep{Gutmark_Grinstein_AnnuRev_1999,Mi2010,Azad_2012,Hashiehbaf2020,Tyliszczak_et_all_IJHFF_2022,Wawrzak_et_al_2024} as they tend to enhance the mixing. A thorough review of twentieth-century research on non-circular jets, with emphasis on their influence on mixing at both large and small scales in low- and high-speed flows, was provided by~\cite{Gutmark_Grinstein_AnnuRev_1999}. More recent investigations have shifted toward studying jets emerging from shaped orifices or from nozzles with smooth contractions, focusing on flow asymmetries, enhancement of directional mixing, variations in velocity decay, the role of nozzle aspect ratio in axis-switching, and the dependence of these features on inlet conditions such as Reynolds number and turbulence intensity.

Examining centreline mixing properties, Mi et al.~\cite{Mi2010,Mi_Nathan2000} compared jets from nine nozzle geometries: one circular nozzle with a smooth contraction (used as a reference) and eight sharp-edged orifices of various shapes, including elliptical, triangular (isosceles and equilateral), square, rectangular, cross-shaped, and five-armed star sections. Their results showed that breaking the initial axisymmetry of the jet enhanced mean axial velocity decay and fluctuation growth, with the isosceles triangle producing the strongest effect.
This outcome contrasts to some degree with Quinn~\cite{QuinnExp_Fluids_2005}, who found that the isosceles triangle led to the fastest velocity decay only in the far field, whereas in the near field an equilateral triangle nozzle produced the shortest potential core and highest decay rate.
Overall, jets from sharp-edged orifices are found to be more energetic than those from smoothly contoured nozzles, leading to a faster axial velocity decay \citep{Mi2010,Azad_2012,Mi_Nathan2000,Quinn_EJM_2006,Deo_2007,Quinn_EJM_2007}.

\begin{figure*}
\centering
\includegraphics[width=0.8\linewidth]{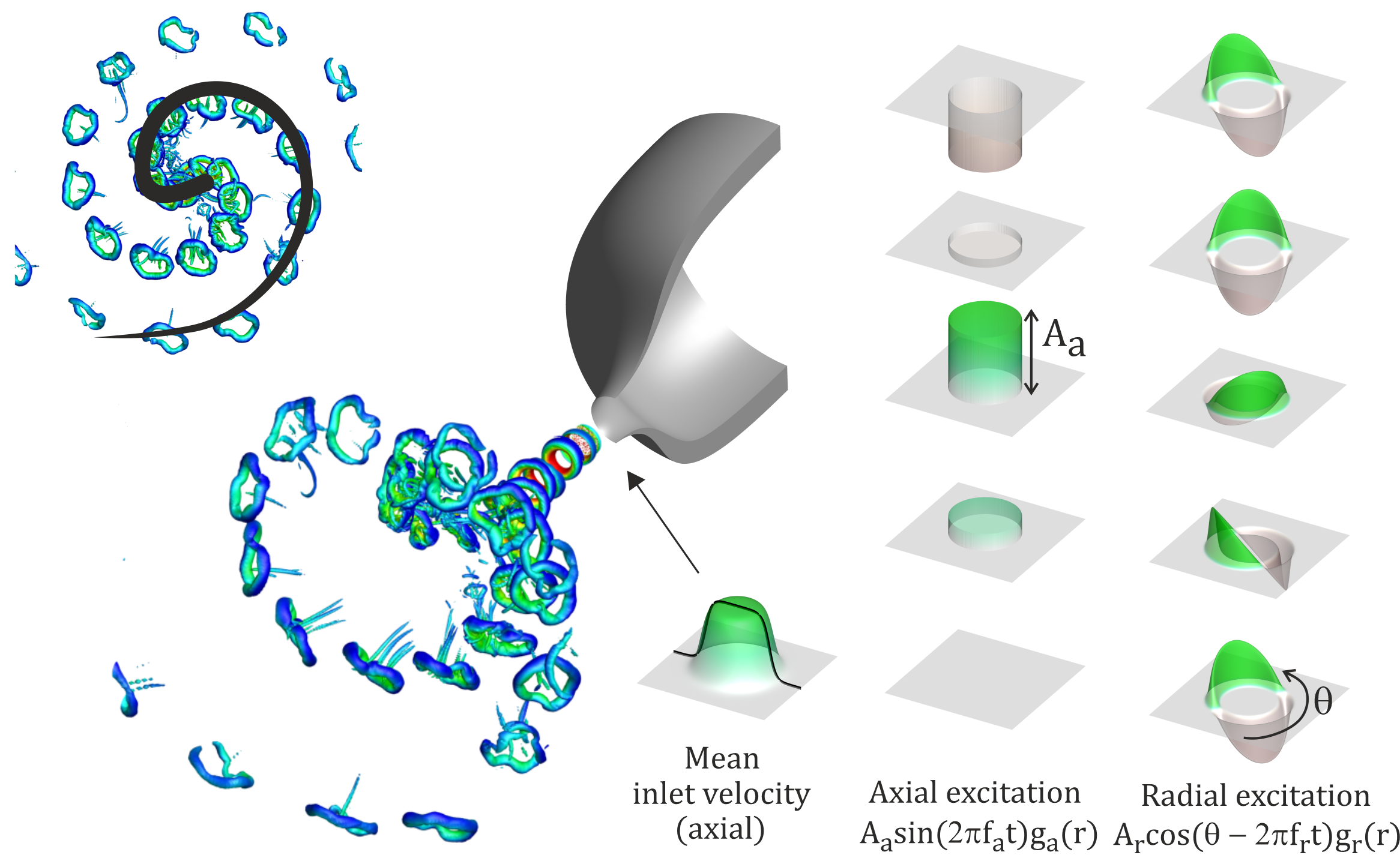}
\caption{Coherent vortex structures in the excited jet with the temporal evolution of axial and radial velocity excitation. }
\label{fig:scheme}
\end{figure*}

In general, PFC techniques are easier to design and manufacture, and less expensive to maintain than active flow control (AFC) methods, making them more feasible for real-world applications. However, PFC is often effective only within a limited range of operating conditions near a flow regime for which they have been optimized. When the operating point significantly diverges, system performance may even degrade. AFC methods require an energy input and involve various types of unsteady actuators, such as piezoelectric devices, magnetic flaps, loudspeakers, mini-jets, and suction slots~\citep{Collis_ProgAeroSci_2004,Cattafesta_AnnRev_2011}. When operated in closed-loop (interactive) mode, AFC offers the flexibility to adapt to changing flow conditions.

Research on AFC of jets dates back to the 1970s with the pioneering experiment of Crow and Champagne~\cite{CrowChampagne_JFM_1971}, who demonstrated that axial excitation at suitable frequencies produces enhanced mixing and velocity fluctuation features absent in natural jets. Their work sparked extensive experimental efforts that confirmed the strong potential of active control \citep{ZamanHussain_JFM_1980,HussainZaman_JFM_1980,ZamanHussain_JFM_1981,ArbeyFfowcsWilliams_JFM_1984,Cho_Exp_in_Fluids_1998,DrobniakKlajny_JoT_2002,Suzuki_et_all_ExpFluids_2004,Balarac_PoF_2007}. More recently, Kanharajau et al.~\cite{Kanharajau_JFM_2020} studied acoustically excited high-speed jets, showing strong interactions between toroidal vortex rings and streamwise vortices, leading to undulations of the ring cores. Typically, AFC employs mass flow excitation via loudspeakers upstream of the nozzle \citep{Chao_ExpFluids_2000,Damare_CaF_2004,Baillot_CST_2007}, often in combination with radial or azimuthal forcing from actuators or synthetic jets \citep{Chao_ExpFluids_2000,Suzuki_et_all_ExpFluids_2004,Kurimoto_ExpFluids_2005,Saiki_CaF_2011}. Such multimode forcing, with tunable amplitude and frequency, can modify not only the mean flow but also the overall jet structure. Spectacular examples of applying AFC to jet flows are bifurcating and blooming jets~\citep{SilvaMetais_PoF_2002,ReynoldsParekh_AnnRev_2003,TyliszczakGeurts_FTaC_2014,Tyliszczak_PoF_2015,Gohil_JFM_2015} characterized by vortical rings disjoining from the main jet stream and flowing in multiple radial directions. A common problem shared by passive and active approaches is to find an optimal design or optimal excitation parameters as this often involves searching over a multiple parameter space. 

Among various optimization methods, such as gradient-based and gradient-free methods, adjoint methods, and genetic and particle swarm optimization algorithms~\citep{Mohammadi_Pironneau_AnnRev_2004,Gad_ArchCompMethEng_2022,Selvarajan_ArtIntellRev_2024}, machine learning (ML) approach~\citep{Alzubaidi_J_Big_Data_2021,Sarker_SN_Computer_Science_2021} seems to be the fastest-developing, as indicated by the growing number of publications each year. Driven by unprecedented volumes of data from experiments, field measurements, and large-scale simulations, ML has emerged as a powerful tool for solving fluid flow problems~\citep{BurtonNoackKoumoutsakos_AnnRev_2020,Vinuesa_Burton_NatCompSci_2022}. ML offers a modular and flexible modeling framework that effectively addresses various challenges, including turbulence closure modeling~\citep{duraisamy2019anfm, ling2016jfm, xiao2016jcp, charalampopoulos2022prf}, reduced-order modeling~\citep{Kaiser2014jfm,Milano20021jcp,Nair2015jfm,GopalakrishnanMeena2018pre}, flow cleansing~\citep{Fukami2019jfm,Lee2017ef,Liang2003ef}, aerodynamic optimization~\citep{Kern2004nc,vanRees2015jfm,Strom2016ne,Hamdaoui2010ja}, and flow control~\citep{blanchard2021ams,Benard2016ef,duriez2017book,colabrese2017prl}. In AFC of jets, ML reveals itself as a very efficient optimization tool for mixing enhancement~\citep{KoumoustakosFreund_AIAA_2001,HilgersBoersma_FDR_2001,Zhou_et_al-JFM-2020,blanchard2021ams,Reumschussel_PoF-2024,Jiang-PoF-2024,Shaqarin_et_al_SciRep_2024, Li_et_al-JFM-2024}. By exploring a large, control parameter space, ML tests seemingly incorrect or counter-intuitive parameter combinations or their values, sometimes resulting in unexpectedly high control performance or the discovery of new phenomena. An example of this situation can be found in a recent work of Li et al.~\cite{Li_et_al-JFM-2024}. This research aimed to determine excitation parameters that ensure the most uniform velocity distribution at a specified distance from the jet nozzle exit by combining Bayesian optimization (BO), deep learning (DL), and persistent data topology. Among 1000 ML-driven cases, the optimal BO-DL and BO solutions proved particularly interesting. The former exhibited the blooming multiarmed jet topology already known in literature ~\citep{ReynoldsParekh_AnnRev_2003,Tyliszczak_PoF_2015,Gohil_JFM_2015,Tyliszczak_IJHFF_2018}, achieved, however, at significantly lower excitation energy cost than reported so far. The BO solution, on the other hand, revealed a new jet pattern with helix-type spiral arms formed by toroidal vortical structures, as shown in Fig.~\ref{fig:scheme}. 
Given decades of research on jet dynamics, discovering previously unknown aspects of its spatio-temporal evolution is especially noteworthy - not only from a scientific perspective but also for its practical significance. It turns out that this type of excitation enhances mixing to an unprecedented degree compared to that observed in any previously known jet topologies. In~\cite{Li_et_al-JFM-2024}, the occurrence of the spiral pattern was interpreted as the result of the rotation of two bifurcating jet branches. This paper demonstrates that this interpretation is incomplete and convincingly explains the true mechanism behind the spiral formation. Moreover, we show that the helix-jet with two pseudo-rotating spirals is not the only solution of this type. We formulate excitation rules enabling the generation of multi-spiral jets with 5, 7 or more spirals.  The proposed control methodology opens new possibilities for various applications, such as combustion devices or heat exchangers, where jet dynamics and patterns enhancing the mixing process are key performance factors.


\section{Methods}
\subsection{Mathematical model}

The research is performed by applying the large eddy simulation (LES) method~\citep{SagautBook,GeurtsBook}. The LES method has been developing for more than 60 years. Extensive research on sub-grid modelling techniques, suitable discretization methods, commutation errors and mutual error interactions carried out over this period led to the current maturity of the LES method and its perception as a reliable tool in simulations of turbulent flows involving complex physical processes (laminar-turbulent transition, hydrodynamic instability, two-phase problems, solid-fluid interactions, reacting flows, etc.)~\citep{Georgiadis2010, Zhiyin2015, Geurts2019}.

For incompressible, constant-density flows the continuity and the Navier–Stokes
equations in the framework of LES are given as

\begin{equation}
\frac{\partial \bar{U}_i}{\partial x_i}=0
\end{equation}
\begin{equation}
\frac{\partial \bar{U}_i}{\partial t} + \frac{\partial \bar{U}_i\bar{U}_j}{\partial x_j} =-\frac{1}{\bar{\rho}}\frac{\partial \bar{P}}{\partial x_i}+\frac{\partial}{\partial x_j}\left[ \left( \nu+\nu_{sgs} \right) \left( \frac{\partial \bar{U}_i}{\partial x_j}+\frac{\partial \bar{U}_j}{\partial x_i} \right)  \right]
\end{equation} 

\noindent where the overbar represents spatial filtering. The variables $\bar{U}_i, \bar{P}$ and $\rho$ are the velocity components, pressure and density, respectively. The symbols $\nu$ and $\nu_{sgs}$ are the kinematic viscosity and sub-grid viscosity which in the present work is modelled using the Vremans' sub-grid model~\citep{Vreman_SGS}.

\begin{figure*}
\centering
\includegraphics[width=0.8\linewidth]{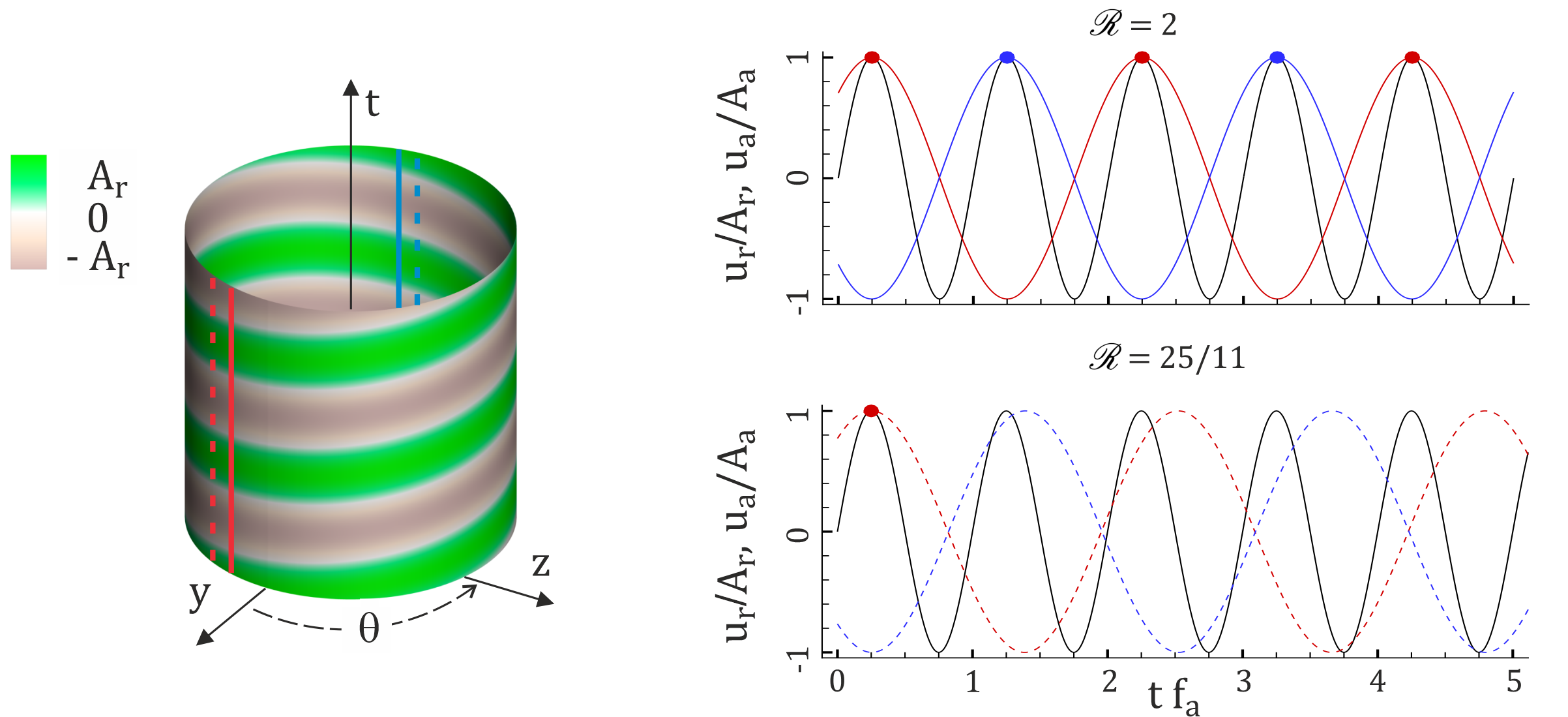}
\caption{Left figure: spatio-temporal variability of the radial excitation $u_r$. The solid (red/blue) and dashed (red/blue) lines indicate the azimuthal locations at which the $u_r$ evolutions are presented in the figures on the right-hand side with corresponding line patterns and colours. The solid black line represents the $u_a$ evolution.}
\label{fig:helix_origin_contours}
\end{figure*}

\subsection{Computational domain and boundary conditions}\label{subsec:domain_and_BC}

The configuration is a jet flow exiting a circular nozzle of diameter $D$ as schematically shown in Fig.~\ref{fig:scheme}. The flow is described in the Cartesian coordinate system $(x,y,z)$ where $x$ represents the streamwise direction and the origin coincides with the center of the nozzle. The computational domain is a rectangular cuboid ($L_x \times L_y \times L_z = 16D \times 12D \times 12D$). The geometry of the jet nozzle is not included in the simulations. Instead, as in several studies on modelling turbulent jet flows~\citep{DanailaBoersma_PoF_2000,SilvaMetais_PoF_2002,Foysi_et_al_IJHFF_2010,Gohil_IJHFF_2015,Gohil_JFM_2015,Li_et_al-JFM-2024}, the flow emerging from the nozzle is represented by instantaneous axial and radial velocity components defined as

\begin{equation}\label{eq:U_a_prof} 
\begin{split} U_a(r, t) &= U_{m}(r) + u_a(r, t) + u_{turb}(r, t), \\
U_r(r, t) &= u_r(r, t). 
\end{split} 
\end{equation}

\noindent where $t$ is the time axis and $r=\sqrt{z^2+y^2}$ is radial coordinate at the nozzle exit. The mean turbulent velocity profile is assumed as tangent hyperbolic function 

\begin{equation}\label{eq:mean-u}
U_{m}(r) = \frac{U+U_c}{2}-\frac{U-U_c}{2}\tanh\left(\frac{1}{4}\frac{R}{\delta}\left(\frac{r}{R}-\frac{R}{r}\right)\right)
\end{equation}

\noindent where $R=D/2$ is a virtual nozzle radius, $U$ - the jet centerline velocity, $U_c=0.03U$ - a co-flow velocity, and $\delta=0.05R$ is the momentum thickness. The terms $u_a$ and $u_r$ are the axial and radial excitation components (see Section~\ref{sec:methodology}), and $u_{turb}$ represents turbulent fluctuations computed by applying a digital filtering method proposed by~\cite{Klein_JCP_2003}. The pressure at the inlet plane is calculated from the Neumann boundary condition, $\nabla p\cdot \mathbf{n}$, with $\mathbf{n}$ the unit vector normal to the boundary. At the side boundaries, the axial velocity is set to be $U_c$, the remaining components are assumed to be zero, and the pressure is calculated using $\nabla p\cdot \mathbf{n}=0$. This boundary condition prevents a natural suction induced by the jet stream. The added co-flowing stream mimics its lack. As shown by~\cite{SilvaMetais_PoF_2002}, the co-flow at a level of $U_c\le 0.1U$ does not change the jet dynamics. At the outflow plane, the pressure equals zero, and the velocity components are computed from a convective boundary condition~\citep{Orlanski_JCP_1976}. It minimizes the impact of limited domain size on the upstream flow and vortices leaving the computational domain.

\section{Methodology}\label{sec:methodology}

In experimental research, excitation typically results from a combination of acoustic waves and mechanical or fluidic forcing. Acoustic waves are generated by loudspeakers positioned upstream of the nozzle exits \citep{Chao_CST_2002,Damare_CaF_2004,Baillot_CST_2007,Abdurakipov_CST_2013}, while mechanical and fluidic forcing is achieved through magnetic or piezoelectric flapping elements~\citep{ParekhLeonard_TF35_1988,Chao_ExpFluids_2000,Suzuki_et_all_ExpFluids_2004,Kurimoto_ExpFluids_2005,Saiki_CaF_2011} and mini-jets~\citep{Zhou_et_al-JFM-2020,Reumschussel_PoF-2024,Jiang-PoF-2024,Shaqarin_et_al_SciRep_2024} mounted at the nozzle lips. These techniques enable the generation of various types of excitation, including flapping and azimuthally travelling disturbances, while acoustic excitation induces mass flow oscillations. 
The present research employs both types of excitation by imposing velocity fluctuations on the main jet stream ($u_a$ and $u_r$ in~\eqref{eq:U_a_prof}). \cite{Li_et_al-JFM-2024} applied LES in combination with BO and BO-DL methods to identify optimal settings within a 22-dimensional vector of control parameters, including the amplitudes and oscillation frequencies of $u_a$ and $u_r$. Based on a sensitivity analysis of the BO and BO-DL optimal solutions resulting in multi-armed and helical jets as reported in \cite{Li_et_al-JFM-2024}, we propose a particularly simple excitation form capable of generating both jet types, while also enabling theoretical analysis. It is defined as

\begin{equation}
\begin{split}
    u_a(r,t)&=A_a\sin({2\pi f_a t})g_a(r)\\  u_r(r,t)&=A_r\cos({\theta - 2\pi f_r t})g_r(r)
\end{split}
    \label{eq:excitation}
\end{equation}

\noindent where $\theta=\arctan(z/y)$ is the azimuthal coordinate at the nozzle exit. The symbols $A_a$, $A_r$ are the excitation amplitudes, $f_a$, $f_r$ stand for the excitation frequencies, and $g_a(r)$, $g_r(r)$ are the masking functions, which limit the radial extent of the excitation. They are defined as 

\begin{equation}
\begin{split}
    g_a(r) &= \begin{cases}
1, & \text{for} \,\, r \leq R, \\
0, & \text{otherwise}
\end{cases} \\ 
g_r(r)&=\exp\left(-1000(|R-r|)^{2.5}\right)
\end{split}
\end{equation}

The former confines $u_a$ to the nozzle area, while the latter limits $u_r$ to a thin shear layer region. 
The applied excitation reflects real-world scenarios by mimicking the excitation induced by a loudspeaker ($u_a$) and the one generated by magnetic or piezoelectric flapping elements mounted on the nozzle edge ($u_r$).

\begin{figure*}
\centering
\includegraphics[width=0.8\linewidth]{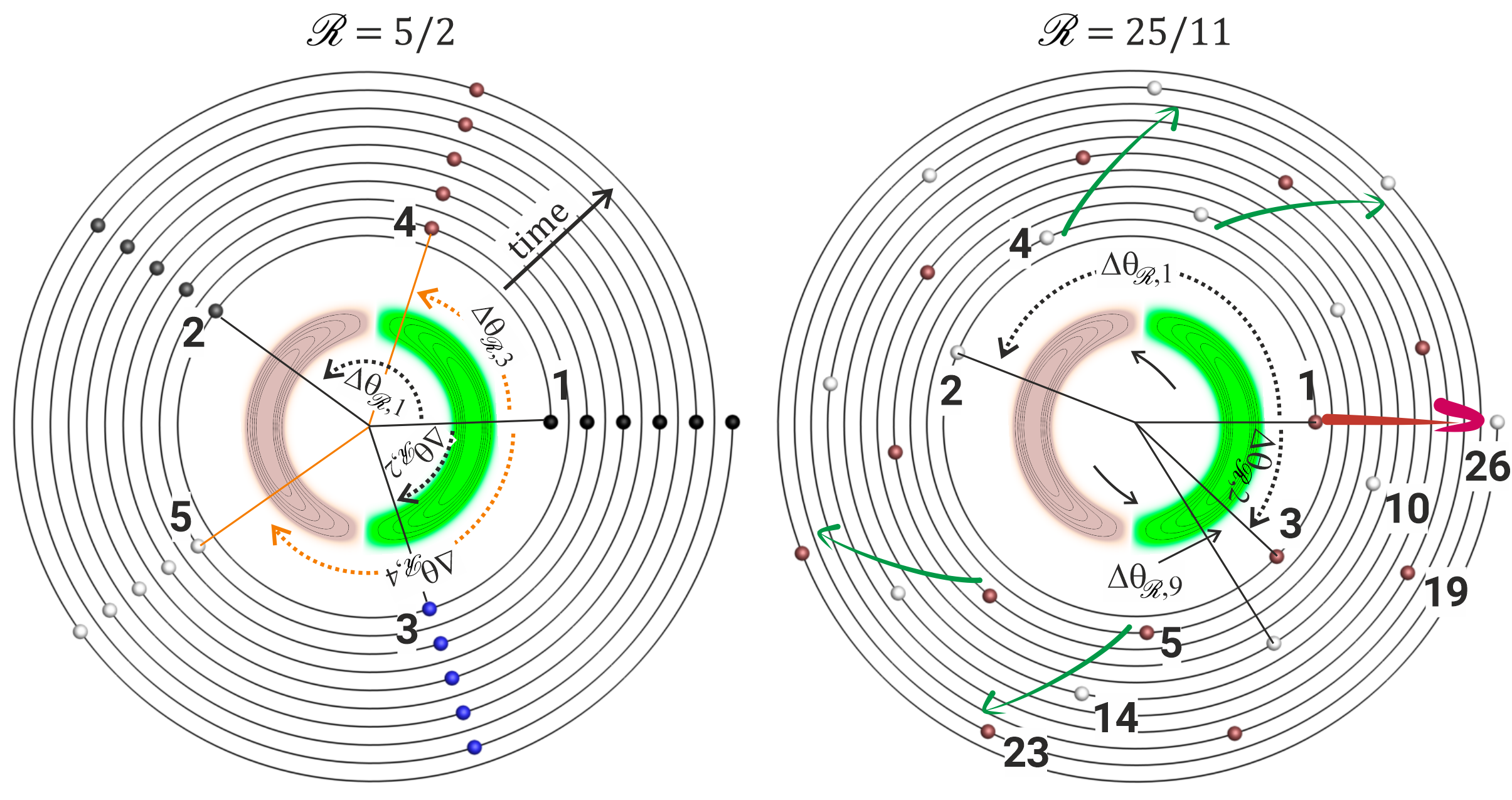}
\caption{Schematic representation of a spatio-temporal distribution of JEPs represented by colour spheres for $\mathcal{R}=5/2$ (left figure) and $\mathcal{R}=25/11$ (right figure). Red and green arrows denote possible combinations of JEPs leading to straight and curved arms forming. Green and light brown contours show $u_r$ distribution at the moment of the first JEP occurrence.}
\label{fig:helix_origin_scheme}
\end{figure*}

Figure~\ref{fig:scheme} illustrates the temporal evolution of $u_a$ and $u_r$ during a single excitation cycle. The mass flow oscillation induced by $u_a$ generates toroidal vortices seen in the nozzle vicinity in Fig.~\ref{fig:scheme}. Simultaneously, $u_r$ acts on the shear layer as if aiming to displace the vortices off-axis. When $u_r$ is positive, the shear layer is pulled outside the main jet stream, and when $u_r$ is negative, it is pushed towards the jet axis. The amplitudes $A_{a,r}$ determine the excitation energy. As shown by~\cite{TyliszczakGeurts_FTaC_2014}, their levels should be comparable to or greater than the turbulence intensity at the nozzle exit for the excitation to be effective. If this is fulfilled, the factor deciding on the jet shape is the ratio of the excitation frequencies $\mathcal{R}=f_a/f_r$, which in terms of the Strouhal numbers, ${St}_a=f_aD/U$, ${St}_r=f_rD/U$, is equivalent to $\mathcal{R}={St}_a/{St}_r$. For $\mathcal{R}=2$ or $\mathcal{R}=3$, bi-furcating and tri-furcating jets appear with the jet core divided into 2 and 3 separate branches, as confirmed in many papers, e.g.~\cite{SilvaMetais_PoF_2002,Tyliszczak_IJHFF_2018}. For $\mathcal{R}$ being non-integer, it has been shown that for specific values of $\mathcal{R}=m/n$ ($m$, $n$ - integer numbers), e.g., $\mathcal{R}=m/n=5/2$, $\mathcal{R}=7/3$ or $\mathcal{R}=13/5$, the vortex rings follow precisely defined paths.
In this case, the multi-armed jets arise with $m$ arms spaced in the azimuthal direction by $\Delta\theta_{\mathcal{R}}=2\pi/m$.
In general, in the range $2<\mathcal{R}\le3$ the existence of 3, 4, 5, 7, 8, 9, 11, 12, 13 and 20-armed jets has been demonstrated~\citep{Tyliszczak_PoF_2015,Gohil_JFM_2015,Tyliszczak_IJHFF_2018}. The intriguing question is "\textit{How does the jet shape evolve when $\mathcal{R}$ is either an irrational number or rational with large $m$?}" The BO-optimization procedure of~\cite{Li_et_al-JFM-2024} did not make any assumption on $\mathcal{R}$ and, as previously mentioned, led to the optimal solution with two arms forming the helix and looking as if they were rotating around the jet axis. 

\section{Simulation details}
The research is conducted for a Reynolds number $Re = UD/\nu$ equal to 3000 following~\cite{Tyliszczak_IJHFF_2018} and~\cite{Li_et_al-JFM-2024}. Using a relatively low $Re$ allows for obtaining accurate solutions with a relatively coarse computational mesh. The excitation amplitudes are taken $A_a=A_r=0.08U$ and the Strouhal number of the axial excitation is fixed and assumed $St_a=0.5$. It fits in the range of the Strouhal number of the preferred mode at $St_p= 0.3-0.64$~\citep{CrowChampagne_JFM_1971,GutmarkHo_PoF_1983,Sadeghi_Pollard_PoF_2012} and in the range of $St_a=0.15-0.8$ for which the bifurcating and blooming jets are observed~\citep{LeeReynolds_TF22_1985,ReynoldsParekh_AnnRev_2003,Tyliszczak_IJHFF_2018}. The turbulent fluctuations $u_{turb}$ are assumed at a level of $0.01U$. Their impact on the flow dynamics is very small compared to the impact of the excitation with $A_a=A_r=0.08U$. Nevertheless, we include it to reflect the real situation where some turbulent velocity fluctuations are unavoidable.

\begin{figure*}[htbp!]
\centering
\includegraphics[width=0.4\linewidth]{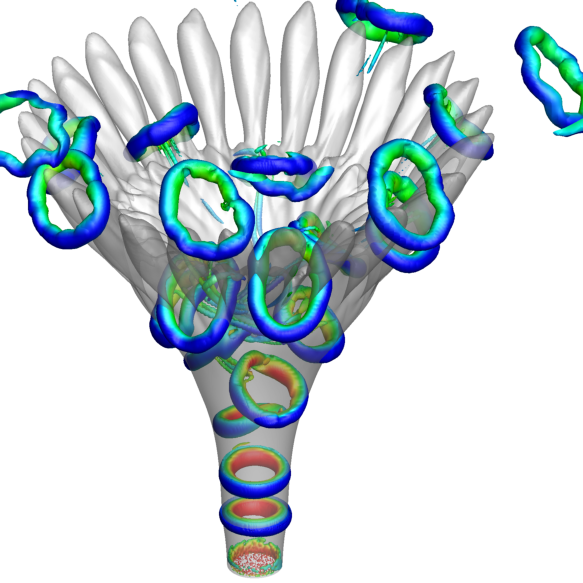}
\includegraphics[width=0.4\linewidth]{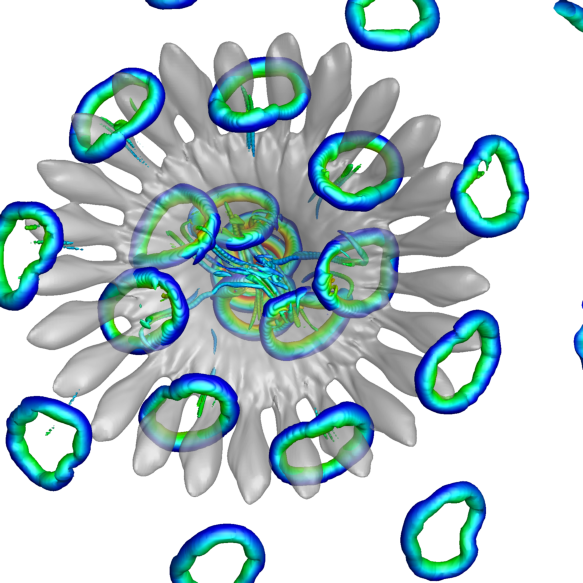}
\caption{Isosurface of the Q-parameter ($Q=1.0 (U/D)^2$) - vortical rings coloured by the vertical velocity component normalised by the inlet jet velocity ($U_x/U$) and the isosurface of its time-averaged value $\langle U_x\rangle/U =0.1$ in the jet excited at $\mathcal{R}=25/11$. View from the side (a) and top (b). }
\label{fig:Q_25_11}
\end{figure*}

\section{Results}
\subsection{Origin of spirals}

For ease of presentation, we consider the cases with $\mathcal{R}=2$ and $\mathcal{R}=25/11$. 
Figure~\ref{fig:helix_origin_contours} illustrates the evolution of the velocity excitation in a spatio-temporal coordinate system. 
The cylindrical surface shows the variability of $u_r$ at the nozzle radius $r=D/2$. 
The excitation varies within the $[-A_r, A_r]$ range over space and time, completing a full cycle in time $t_r = 1/f_r$. At any moment, there are two azimuthal locations where $u_r$ is maximum and minimum, $\theta=2\pi f_rt$ and $\theta=2\pi f_rt + \pi$. 
The evolution of $u_r$ along the red lines placed at the locations denoted $\theta_{\mathcal{R}=2}$, $\theta_{\mathcal{R}=25/11}$, and the blue lines at $\theta=\theta_{\mathcal{R}} + \pi$ is presented in the figure aside. The colour and pattern of the lines (solid, dashed) reflect $u_r$ at the particular $\theta$ angles. The black solid line represents the axial excitation ($u_a$), which level varies in time but is independent of $\theta$. It can be observed that there are time instances when both $u_r$ and $u_a$ simultaneously attain their maximum values. These joint excitation peaks, hereafter called JEPs, are marked with red and blue circles. Taking into account the formulas defined in equation~(\ref{eq:excitation})
one can easily infer that the first JEPs appear at the time moment $t_{JEP,1}=1/(4f_a)$ in the locations $\theta_{\mathcal{R},1}=\pi/(2\mathcal{R})$, which for assumed $\mathcal{R}$ equals $\theta_{\mathcal{R}=2,1}={\pi}/{4}$ and $\theta_{\mathcal{R}=25/11,1}={11\pi}/{50}$. When time passes, the maximum of $u_r$ moves along $\theta$ and $u_a$ oscillates. For $\mathcal{R}=2$, the successive $k$ JEPs for $k=2,3,\dots,\infty$ occur in the locations $\theta_k=\theta_{\mathcal{R}=2,1}+k\pi$ at time instances $t_{k, JEP}=t_{k,JEP,1}+(k-1)\Delta t_{JEP}$ (blue/red circles), where $\Delta t_{JEP}={1}/{f_a}$. Whenever JEP occurs, the toroidal vortex is generated, which moves downstream, being advected by the main jet stream. The radial disturbance grows and causes a slow inclination of subsequent vortices in alternate directions in the plane $x-\theta_{\mathcal{R}=2}$. Approximately five diameters from the nozzle exit, the vortices become tilted such that the main jet stream splits into two distinct branches. This mechanism leads to the formation of the bifurcating jet~\citep{ReynoldsParekh_AnnRev_2003}. 

The situation for ${\mathcal{R}=25/11}$, which does not differ significantly from ${\mathcal{R}=2}$, is, however, significantly distinct. As can be seen in Fig.~\ref{fig:helix_origin_contours}, after the occurrence of the first JEP, there is a clear mismatch between the subsequent $u_a$ and $u_r$ maxima. This, however, does not imply that JEP cannot occur in locations other than $\theta=\theta_{\mathcal{R}=25/11,1} + \pi$. Taking into account equation~(\ref{eq:excitation}) and knowing that the next maximum of $u_a$ occurs at $t={1}/({4f_a})+{1}/{f_a}$, it can be shown that for ${\mathcal{R}=25/11}$ the 2nd JEP will appear at $\theta = {11\pi}/{10}$. Hence, the shift between the 2nd and 1st JEP at $\theta = {11\pi}/{50}$ is $\Delta\theta={22\pi}/{25}$, which means that subsequent JEPs are not in a plane. Continuing this analysis, it can be shown that, in general, JEPs occur at

\begin{equation}
    \theta_{\mathcal{R},k}=\frac{\pi}{2\mathcal{R}}(4k-3)
    \label{eq:theta_rk}
\end{equation}

\noindent For $\mathcal{R}$ being a rational number, the $\theta_k$ locations are finite, meaning that at some moment, JEP must occur in its initial location. This $k$-th JEP can be determined from the following relation 

\begin{equation}
    \theta_{\mathcal{R},k} -(k-1)\frac{2\pi}{\mathcal{R}} = \theta_{\mathcal{R},1}. 
\end{equation}

\noindent when $m$ and $n$ defining $\mathcal{R}=m/n$ are known then $k=m+1$, which happens after $n$ full rotations of the $u_r$ maximum. Thus, the number of JEPs, which occur along the azimuthal direction equals $m$ and the time moments when JEPs appear in these $m$ locations are spaced by $\Delta t_{\mathcal{R}}=m/f_a$. 

As in the case of the bifurcating jet with the arms formed in $x-\theta_{\mathcal{R}=2}$ plane, for ${\mathcal{R}=25/11}$ the generated vortices will incline and tend to separate into 25 vortical paths on $x-r$ half-planes at $\theta_{\mathcal{R},1-25}$. 
The occurrence of the vortices, however, does not necessarily imply splitting into easily detectable arms. This depends on a sequence where JEPs occur in all $\theta_m$ locations. The azimuthal distance between two arbitrary JEPs located at $\theta_k$ and $\theta_{k \pm l}$ is given by

\begin{equation}
\Delta \theta_{\mathcal{R},l} = |\theta_{k\pm l} -\theta_k|= \min \left( \operatorname{mod}_{2\pi} \left( l \frac{2\pi}{\mathcal{R}} \right), \, 2\pi - \operatorname{mod}_{2\pi} \left( l \frac{2\pi}{\mathcal{R}} \right) \right).
\label{eq:delta_theta}
\end{equation}

\begin{figure*}[htbp!]
\centering
\includegraphics[width=0.8\linewidth]{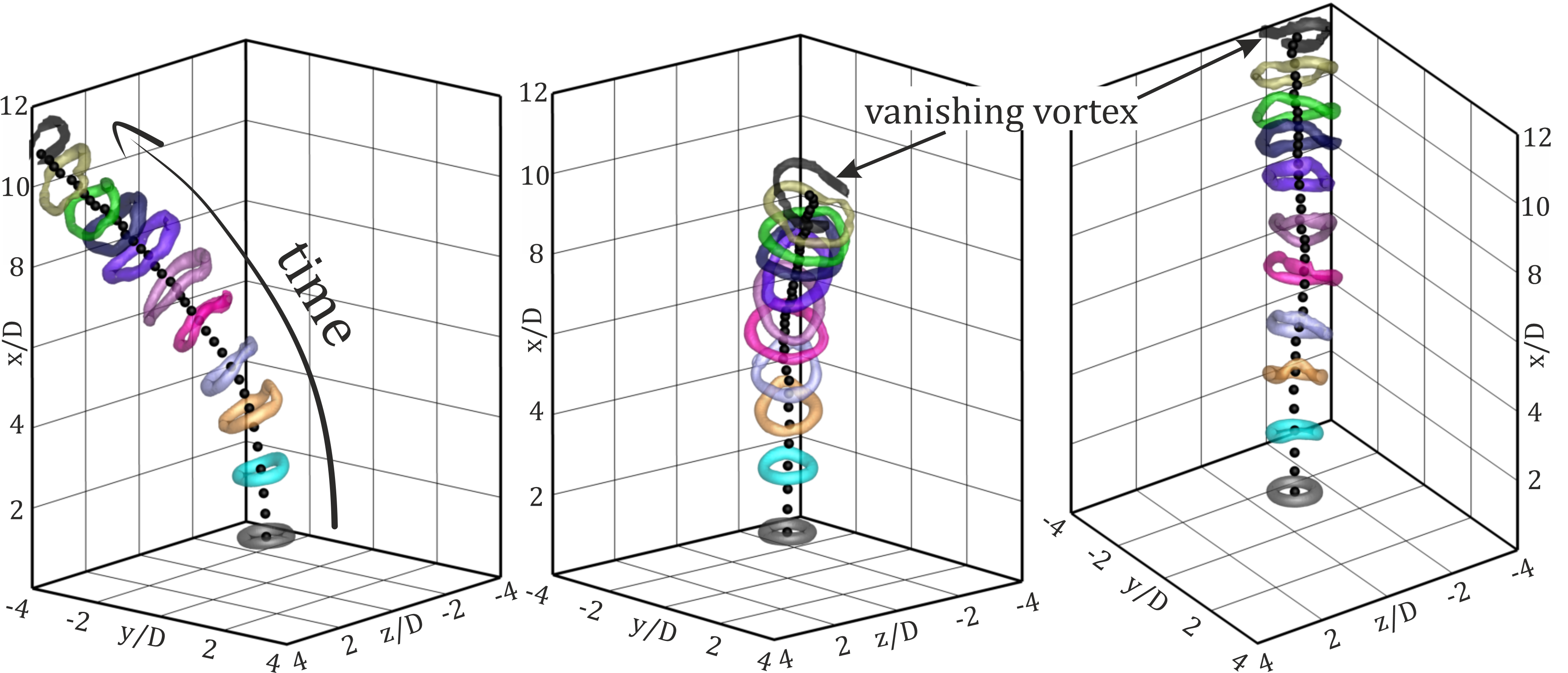}
\caption{ 
Temporal evolution of an isolated vortex ring ($Q$-parameter isosurface, $Q=1.0 (U/D)^2$) seen from different sides. Colours facilitate immediate identification of the vortex location in particular subfigures. Black spheres represent the centres of mass of the fluid enclosed by $Q=1.0 (U/D)^2$. Results for the jet excited at $\mathcal{R}=25/11$.}
\label{fig:vortices_path_single}
\end{figure*}

For instance, $\Delta\theta_{\mathcal{R},2}$ and $\Delta\theta_{\mathcal{R},5}$ denote distances between every 2nd and every 5th JEP when they appear in time. Note that for $l=m$, $l=2m$, etc., $\Delta \theta_{\mathcal{R},l}=0$, which means that JEP occurs in the initial location $\theta_{\mathcal{R},1}$. Also, note that $\Delta \theta_{\mathcal{R}, l} \ne l \Delta \theta_{\mathcal{R}}$. This difference is illustrated in Fig.~\ref{fig:helix_origin_scheme}, where spheres denote the locations of subsequent JEPs (1,2,3,...) and the black spiral line refers to time, i.e., the further the sphere is located on this spiral the later the JEP is generated. The green and light brown contours show the distribution of the $u_r$ excitation at the moment of the first JEP. When looking at the nozzle from the top, $u_r$ rotates counter-clockwise. For $\mathcal{R}=5/2$ for which $\Delta \theta_{\mathcal{R}}=2\pi/5 = 72^{\footnotesize \textrm{o}}$, we have $\Delta \theta_{\mathcal{R},1}=144^{\footnotesize \textrm{o}}$, $\Delta \theta_{\mathcal{R},2}=72^{\footnotesize \textrm{o}}$, $\Delta \theta_{\mathcal{R},3}=72^{\footnotesize \textrm{o}}$, $\Delta \theta_{\mathcal{R},4}=144^{\footnotesize \textrm{o}}$, $\Delta \theta_{\mathcal{R},5}=0^{\footnotesize \textrm{o}}$, whereas for $\mathcal{R}=25/11$ with $\Delta \theta_{\mathcal{R}}=2\pi/25=14.4^{\footnotesize \textrm{o}}$, we have $\Delta \theta_{\mathcal{R},1}=158.4^{\footnotesize \textrm{o}}$, $\Delta \theta_{\mathcal{R},2}=43.2^{\footnotesize \textrm{o}}$, $\Delta \theta_{\mathcal{R},3}=115.2^{\footnotesize \textrm{o}}$, $\dots$,  $\Delta \theta_{\mathcal{R},23}=43.2^{\footnotesize \textrm{o}}$, $\Delta \theta_{\mathcal{R},24}=158.4^{\footnotesize \textrm{o}}$ and $\Delta \theta_{\mathcal{R},25}=0^{\footnotesize \textrm{o}}$. From Fig.~\ref{fig:helix_origin_scheme}, it can be easily inferred that for $\mathcal{R}=5/2$ the jet will separate into 5 arms following JEPs locations. For $\mathcal{R}=25/11$, the situation is not obvious. In a spatiotemporal coordinate system, JEPs no. 1, 3, 5, etc. and 2, 4, 6, etc. align along two spiral paths. On the other hand, JEP no. 1 and 26 are in the same plane $x-r$ at $\theta_{\mathcal{R},1}$ and form a pair, as indicated by the red arrow. Similar pairs are for JEPs, 2 and 27, 3 and 28, etc. Additionally, the third jet pattern emerges, formed by JEPs along green arrows, e.g., JEPs no. 5, 14, 23. In this case, nine distinct curved paths can be identified.
So, one may ask: \textit{What would be observed when examining a 'frozen' instantaneous distribution of the vortices?} In theory, observing any of the above jet forms is equally probable and depends on the observer's subjective perception. Figure~\ref{fig:Q_25_11} shows an isosurface of the $Q$-parameter coloured by the vertical velocity component and the grey isosurface representing its time-averaged value $\langle U_x\rangle$. Without showing the latter, identifying the vortex paths would be hardly possible. When viewed from the side, they appear chaotic and randomly distributed. However, when observed from above, one might get the impression that the vortices are aligned along two spiral paths and move along them. Based on the JEP distribution shown in Fig.~\ref{fig:helix_origin_scheme}, one can easily identify that these spiral paths originate from JEPs 1, 3, 5, etc., and 2, 4, 6, etc. The corresponding vortices are generated frequently ($\Delta t_{\mathcal{R},2}=2/f_a$) and relatively close to each other ($\Delta \theta_{\mathcal{R},2}=43.2^\textrm{o}$). The angular speed of JEPs occurring in $\theta_{\mathcal{R},l}$ is defined as 

\begin{equation}\label{eq:omega}
\omega_{\mathcal{R},l}=\Delta \theta_{\mathcal{R},l}/\Delta t_{\mathcal{R},l}.
\end{equation}

Hence, if $\omega_{\mathcal{R},l}$ is large, it means that JEPs occur frequently at intervals of $\Delta t_{\mathcal{R},l}$. Thus, visually connecting the vortices they generate is natural. For instance, the vortices generated by JEPs no. 5 and 14 are much closer to each other in the azimuthal direction ($\Delta \theta_{\mathcal{R},9}=\Delta \theta_{\mathcal{R}}=14.4^\textrm{o}$) than those resulting from JEPs no. 3 and 5. However, the moments when they appear are less frequent ($\Delta t_{\mathcal{R},9}=9/f_a$), which results in $\omega_{\mathcal{R},9}<\omega_{\mathcal{R},2}$. This translates to a large spatial distance between the vortices and the observers' eyes do not connect them. For the same reason, the vortices related to JEPs no. 1 and 26, 2 and 27, etc. 
seem to be uncorrelated. These vortical pairs, however, are located on the 'true' paths revealed by the  $\langle U_x\rangle$ distribution. Up to approximately $8D$ from the nozzle exit, the $\langle U_x\rangle$ isosurface is continuous and resembles the shape of a wine glass with a wavy conical bowl. In this region, the azimuthal distance between the vortices is small ($\Delta \theta_\mathcal{R}=14.4^\textrm{o}$) causing the shear layers to largely overlap. The observed waviness stems from the toroidal shape of the vortices. Downstream, a 25th-finger-like structure forms in the regions where the vortices overlap only partially ($r\Delta \theta_\mathcal{R}\approx D/2$). From Fig.~\ref{fig:Q_25_11}, it can be observed that each vortex engulfs two $\langle U_x\rangle$ fingers, with each finger located in the overlapping region between neighbouring vortices.

In the following discussion, we attempt to quantify spatiotemporal JEPs characteristics and dynamics of vortices in a turbulent flow, affecting the jet pattern perception. Two important properties, which can be deduced from the above $\Delta \theta_{\mathcal{R},l}$ sequences and equation~(\ref{eq:delta_theta}) are the following:

\begin{itemize}
    \item [(1)] Equality of distances $\Delta \theta_{\mathcal{R},l}$ and $\Delta \theta_{\mathcal{R},m-l}$:
{\small   

\begin{flalign}
\Delta \theta_{\mathcal{R},m-l} &= 
   \min \left( 
      \operatorname{mod}_{2\pi}\!\left((m-l)\tfrac{2\pi}{\mathcal{R}}\right),\,
      2\pi - \operatorname{mod}_{2\pi}\!\left((m-l)\tfrac{2\pi}{\mathcal{R}}\right) 
   \right) & \notag \\[6pt]
&=  \min \left( 
      2\pi - \operatorname{mod}_{2\pi}\!\left(l\tfrac{2\pi}{\mathcal{R}}\right),\,
      \operatorname{mod}_{2\pi}\!\left(l\tfrac{2\pi}{\mathcal{R}}\right) 
   \right) & \\[6pt]
&= \Delta \theta_{\mathcal{R}, l} & \notag
\end{flalign}
}
For instance, for $\mathcal{R}=5/2$, $\Delta \theta_{\mathcal{R},1}=\Delta \theta_{\mathcal{R},4}$ and $\Delta \theta_{\mathcal{R},2}=\Delta \theta_{\mathcal{R},3}$, whereas for $\mathcal{R}=25/11$, $\Delta \theta_{\mathcal{R},1}=\Delta \theta_{\mathcal{R},24}$, $\Delta \theta_{\mathcal{R},2}=\Delta \theta_{\mathcal{R},23}$, etc. Hence, it can be deduced that there are only

\begin{equation}
N_{\Delta \theta} =
\begin{cases}
    \frac{m}{2}, & \text{if $m$ is even,} \\
    \frac{m-1}{2}, & \text{if $m$ is odd.}
\end{cases}
\end{equation}

unique $\Delta \theta_{\mathcal{R},l}$ distances.
\item [(2)] The minimum distance $l_{{min}}$ between JEPs indices is determined based on the equation: 
{\small
\begin{flalign}\label{eq:l_lmin}
\Delta \theta_{\mathcal{R},l_{{min}}} = \min \left( \operatorname{mod}_{2\pi} \left( l_{{min}} \frac{2\pi}{\mathcal{R}} \right), \, 2\pi - \operatorname{mod}_{2\pi} \left( l_{{min}} \frac{2\pi}{\mathcal{R}} \right) \right)=\Delta \theta_{\mathcal{R}}
\end{flalign}}

For instance, for ${\mathcal{R}}=5/2$ we find $l_{{min}}=2$, whereas for ${\mathcal{R}}=25/11$,  $l_{{min}}=9$.
\end{itemize}
 The perception of the jet by an observer as multi-armed with $m$ straight arms or with $l_{{min}}$ or $l\ne l_{{min}}$ curved arms depends on the time distances at which the JEPs occur, i.e., $\Delta t_{\mathcal{R}}=m/f_a$, $\Delta t_{\mathcal{R},l_\text{min}}=l_{{min}}/f_a$ and $\Delta t_{\mathcal{R},l}=l/f_a$. Note, that $\Delta\theta_{\mathcal{R},l\ne l_{{min}}}>\Delta\theta_{\mathcal{R},l_{{min}}}$, and hence, $l_{{min}}$ defines the maximum number of curved arms that can be potentially observed.

\begin{figure*}[htbp!]
\centering
\includegraphics[width=0.8\linewidth]{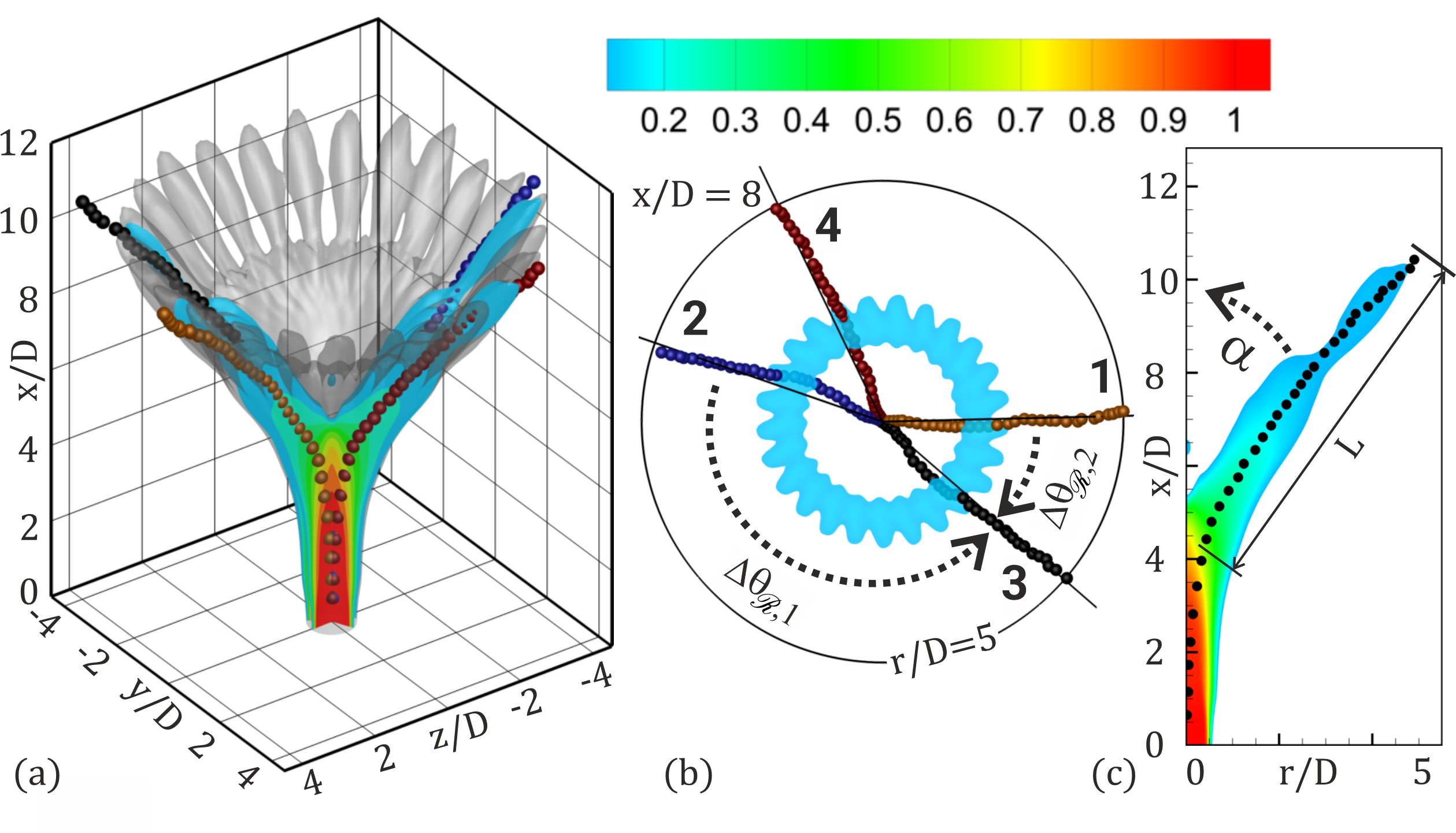}
\caption{Isosurface of the time-averaged vertical velocity component $\langle U_x\rangle/U=0.1$ (a).  Contours of $\langle U_x\rangle/U\ge0.2$ in the cross-sections plane $y-z$ at $x/D=8$ (b) and $r-\theta$ across the finger-like $\langle U_x\rangle$ isosurface (c). Coloured spheres represent the mass centres of the vortices. Results for the jet excited at $\mathcal{R}=25/11$.}
\label{fig:vortices_path}
\end{figure*}

To identify a single straight arm, at least two vortices must simultaneously exist along a radially oriented path. Hence, to observe an $m$-armed jet, $2m$ or more vortices must be present. There are a few factors, which can make the identification of the jet arms impossible. First, the vortices along the path, for convenience called $\mathcal{A}$ and $\mathcal{B}$, cannot be spaced apart from each other by a distance $L_{\mathcal{A}-\mathcal{B}}$ longer than an observable domain ${L}$. Otherwise, $\mathcal{A}$ leaves ${L}$ before $\mathcal{B}$ starts following the same path. This happens when $\mathcal{A}$ and $\mathcal{B}$ are generated too rarely. The condition for the pair of vortices to be simultaneously observed in $L$ is $L_{\mathcal{A}-\mathcal{B}}\le L$. The distance between the vortices can be calculated as $L_{\mathcal{A}-\mathcal{B}}=\Delta t_{\mathcal{R}}U_v$, where $U_v$ is the mean velocity of the vortices along the arm. As shown in \cite{Tyliszczak_PoF_2015,Gohil_JFM_2015,Tyliszczak_IJHFF_2018}, the velocity distribution within the arms closely resembles that observed in typical unexcited jets. In particular, the initial diameters of the arms are comparable with the initial jet diameter $D$. Assuming that the main jet stream fully splits into $m$ arms, the volume flow rate in every arm is $DU/m$. Hence, the initial velocity at the beginning of the arm is $U_a\approx U/m$. The vortices, however, do not move with $U_a$ but with the convection velocity ($U_v$) characteristic for the shear layer, where the vortices are formed. Figure~\ref{fig:vortices_path_single} (see supplementary movie 1) shows an isosurface of the $Q$-parameter for an isolated vortex ring evolving in time for the jet excited with $\mathcal{R}=25/11$. It can be seen that from the distance around $5D$ from the nozzle exit, the vortex moves along an inclined and nearly straight path. Black spheres represent the locations of the centres of mass ($\mathbf{X}_M(t)$) of the fluid enclosed by the $Q$-parameter isosurface.  Knowing $\mathbf{X}_M(t)$, the mean $U_v$ of the vortex along the path can be computed as $U_v=(\mathbf{X}_M(t+\Delta t)-\mathbf{X}_M(t))/\Delta t$. The comparative analysis for various cases shows that $U_v$ weakly depends on $\mathcal{R}$ and can be assumed $U_v=U/4$. Hence, the condition $L_{\mathcal{A}-\mathcal{B}}\le L$ is fulfilled for  

\begin{equation}\label{eq:m_single}
m\le 4 f_a\frac{L}{U} \,\xrightarrow{\,St_a=f_aD\,/\,U\,} \, m\le 4 \,St_a\frac{L}{D}.  
\end{equation}

\begin{figure*}
\centering
\includegraphics[width=0.8\linewidth]{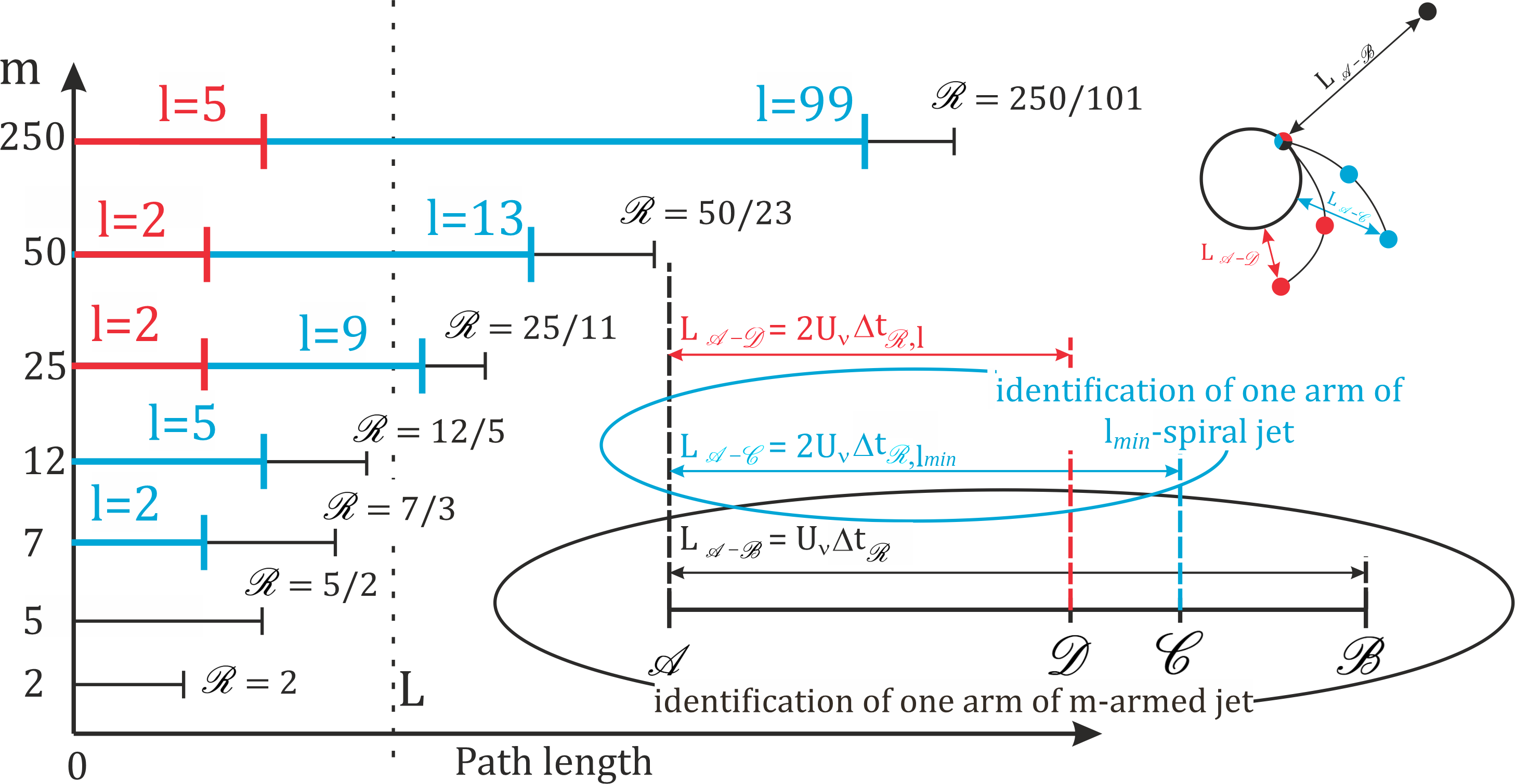}
\caption{Lengths of the vortical paths needed to observe a single straight arm of the $m$-armed jet (black bars) and a single spiral arm (colour bars).}
\label{fig:scheme_L}
\end{figure*}

The vortex path presented in Fig.~\ref{fig:vortices_path_single} can be found for any of the 25 jet arms. Figure~\ref{fig:vortices_path} shows them for the vortices generated by JEPs 1, 2, 3 and 4. It can be seen that the mass centres are not perfectly aligned in the $x-r$  half-planes. The surrounding turbulent flow slightly disturbs their azimuthal locations at particular times. However, assuming these disturbances are not large and 2 vortices suffice to define specific paths, the condition for identifying all $m$ arms of the jet is 

\begin{equation}\label{eq:marmed}  
m_{all}\le 2 \,St_a\frac{L}{D}. 
\end{equation}

\noindent If this condition is not met, the jet pattern likely to be observed is determined by the JEPs spaced by $\Delta \theta_{\mathcal{R},l_{{min}}}$ and occurring every $\Delta t_{\mathcal{R},l_{min}}$. However, based on an example for $\mathcal{R}=25/11$ with $\Delta \theta_{\mathcal{R},l_{{min}}}=\Delta \theta_{\mathcal{R},9}$ (see Figs.~\ref{fig:helix_origin_scheme} and \ref{fig:Q_25_11}) the jet arms generated by JEPs 1-10-19, 3-12-21, etc. could not be identified. If they were, we would observe 9 curved arms, each defined by at least 3 vortical structures. In analogy to the condition allowing for the identification of the straight jet arms one can derive the conditions for the occurrence of the curved arms. The requirement for observing one curved jet arm is

\begin{equation}\label{eq:s_single}
l\le 2 \,St_a\frac{L}{D}  
\end{equation}

\noindent whereas the simultaneous existence of the $l$ curved arms is possible when $l\le l_{all}$ where
\begin{equation}\label{eq:lspir}
l_{all}\le \frac{4}{3} \,St_a\frac{L}{D} .
\end{equation}

\noindent Hence, if $l_\text{min}\le l_{all}$ one should observe $l_\text{min}$ curved arms with the vortices generated every $\Delta t_{\mathcal{R},l_\text{min}}=l_{{min}}/f_a$ and spaced in the azimuthal direction by $\Delta \theta_{\mathcal{R},l_{{min}}}$. However, if $l_\text{min} > l_{all}$ then the observable jet pattern is determined by another JEPs sequence with $l\in N_{\Delta \theta}$ for which (\ref{eq:lspir}) is fulfilled.

The second factor hindering the identification of vortical paths is the lifetime of the vortex, which determines the distance over which the vortex survives in a form that allows its detection. If $L_{\mathcal{A}-\mathcal{B}}\le L$ it is necessary for $\mathcal{A}$ to exist when $\mathcal{B}$ starts following it. 
However, during the vortex motion, its energy dissipates and the surrounding turbulent flow may also deform and destroy the vortex. It can be seen in Fig.~\ref{fig:vortices_path_single} that the vortices are not perfectly toroidal and one located the furthest from the nozzle is fragmented. There is no single method to estimate the vortex lifetime. In general, the lower the flow velocity and turbulence level, the vortices exist for a longer time and travel a greater distance. In~\cite{ReynoldsParekh_AnnRev_2003}, traces of the vortices were observed up to approximately twenty jet diameters from the nozzle exit. 

\begin{figure*}[ht!]
\centering
\includegraphics[width=0.8\linewidth]{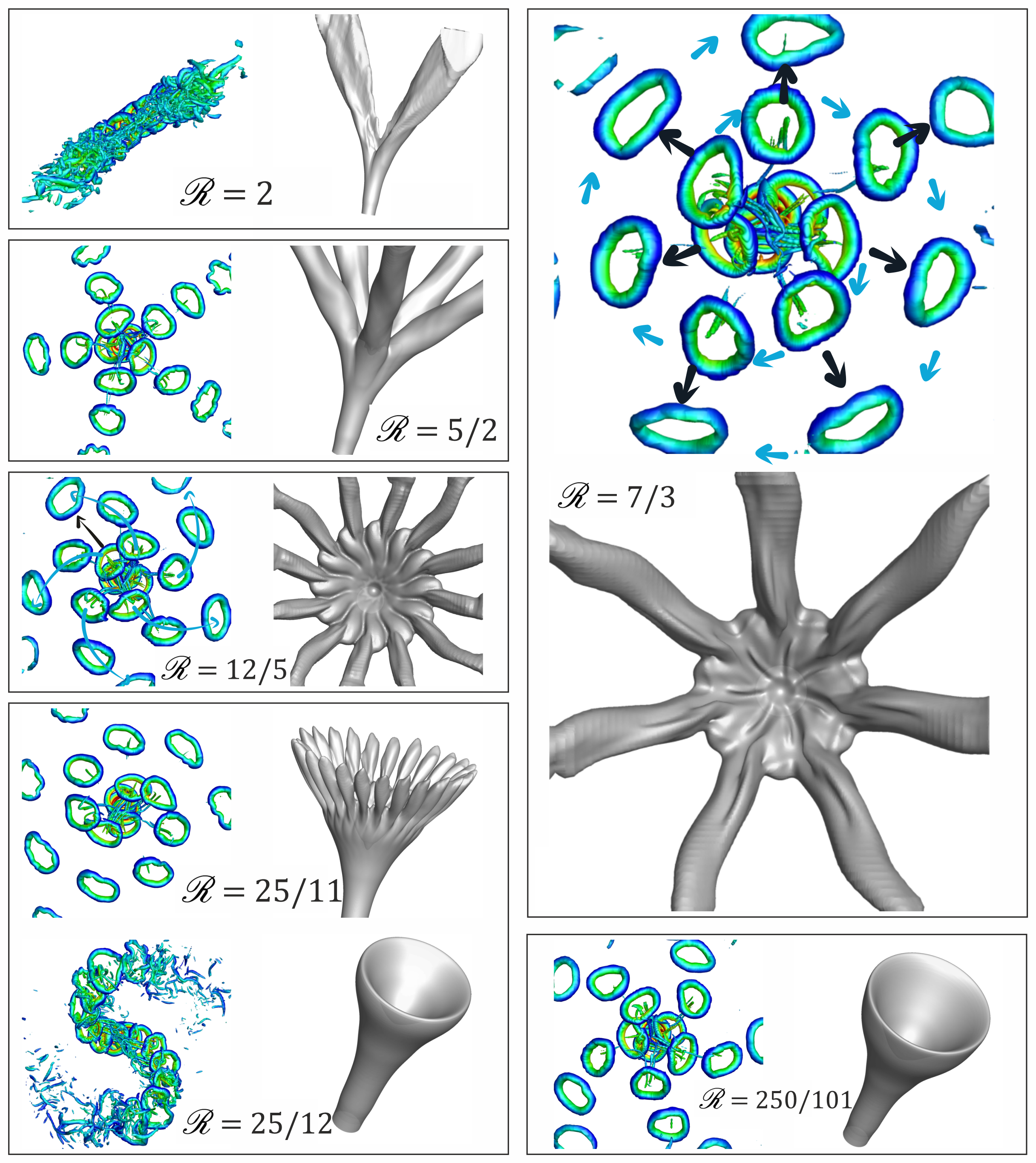}
\caption{Isosurfaces of the Q-parameter ($Q=1.0 (U/D)^2$) coloured by the vertical velocity component
normalised by the inlet jet velocity (Ux/U) and the isosurfaces of the time-averaged value of the vertical velocity component ($\langle U_x\rangle/U =0.1$) in the jets excited with various $\mathcal{R}$.}
\label{fig:helix_delta_n}
\end{figure*}

Finally, it should be noted that a vortex can be destroyed by another vortex if the latter is generated in its immediate vicinity within a short time. This may happen when $\Delta \theta_{\mathcal{R},l}$ is small and the vortices are generated too frequently. In this case, the vortex $\mathcal{A}$ will not `escape' before $\mathcal{B}$ appears in its neighbourhood. In effect, $\mathcal{A}$ and $\mathcal{B}$ mutually interact, deform, and likely destroy. In such a situation distinguishing the separate vortical paths is not a matter of their identification, they do not exist.

\subsection{Specific examples}
Referring to the above discussion, we focus on test cases where the instantaneous jet patterns are univocally classified as multi-armed and those that, despite being multi-armed in reality, are interpreted as spiral. Independently of $\mathcal{R}$, all analysed jets start to split at a distance of approximately $4.5D$ downstream of the nozzle exit. Then, the vortices follow paths inclined to the main jet axis at an angle of $\alpha\approx 40^\textrm{o}$, see Fig.~\ref{fig:vortices_path}(c). The distance from the splitting point to the domain boundary is $9.4D$. However, considering the vortex destruction process, the path length along which the vortices remain well-formed is $L \approx 7.8D$. Figure~\ref{fig:scheme_L} shows a diagram in which horizontal black bars represent the path lengths $L_{\mathcal{A}-\mathcal{B}}$ required to see a single straight arm formed by two or more vortices for various $\mathcal{R}$. It can be seen that only for $\mathcal{R}$ with $m\le12$ the length $L_{\mathcal{A}-\mathcal{B}}<L$ (vertical dashed line) and in these situations, one could point at least one straight arm. The blue bars $L_{\mathcal{A}-\mathcal{C}}$ denote the distance between three vortices generated by JEPs spaced by $l_{min}$ and forming the curved arm. See the inset figure in Fig.~\ref{fig:scheme_L}. This arm would also be visible for the cases with $m\le12$. The red bars $L_{\mathcal{A}-\mathcal{D}}$ represent the paths shorter than $L$ and generated by JEPs spaced by $l<l_{min}$. In these cases $m$ can be large. 

Referring to equation (\ref{eq:m_single}), for $St_a=0.5$ the maximum $m$ for which a single arm can be identified is $m \leq 15.8 \xrightarrow{{\small\textrm{INT}}} 15$, whereas from equation (\ref{eq:marmed}) it follows that 
the maximum number of arms, which can be simultaneously observed equals $m\leq m_{all}= 7.8\to 7$. If the assumed $\mathcal{R}$ is such that $m > m_{all}$, the jet pattern likely to be observed is determined by $l$ curved arms. According to (\ref{eq:s_single}), to see one curved arm, it is necessary that $l\le 7.8\to 7$, while all arms will be simultaneously visible when $l \leq l_{all} = 5.2 \to 5$ (see equation~(\ref{eq:lspir})). 

Figure~\ref{fig:helix_delta_n} shows a set of figures presenting isosurfaces of the $Q$-parameter coloured by the vertical velocity component, and the grey isosurfaces representing its time-averaged value $\langle U_x\rangle$ obtained for various $\mathcal{R}$. 
For $\mathcal{R}=2$ and $\mathcal{R}=5/2$ for which $m\le m_{all}$ two and five well-defined arms are seen both in the instantaneous and time-averaged solutions. Particularly interesting is the case for $\mathcal{R}=7/3$ for which both conditions are fulfilled, i.e., $m\le m_{all}$ and $l_{min}\le l_{all}$. The $\langle U_x\rangle$ isosurface reveals the 7-armed jet pattern, however, when observing the instantaneous solution the situation is ambiguous. The black arrows indicate the paths of the vortices flowing along seven straight arms. In contrast, the blue arrows show the vortices aligning along two curved paths. As a result, they form two long spiral arms. This raises the question: \textit{What would an observer see if the arrows were not drawn - the spirals or the straight arms?} Answering this question is difficult as both responses seem equally probable and depend on individual observer perception. The situation for $\mathcal{R}=12/5$ is much easier, as for this case only the condition $l_{min}\le l_{all}$ is fulfilled. In this case, if the black arrow were not drawn, identifying a single jet arm would require particular attention and could not be regarded as a certainty. In contrast, five curved arms show out almost immediately and mask the true twelve-armed jet pattern. 

\begin{table*}[ht!]
\centering
\caption{\label{tab:JEP_param} Parameters of JEPs for various $\mathcal{R}$. The symbol $l_{int}$ denotes the increment of JEPs indices spaced by $\Delta t^*_{\mathcal{R}, l_{int}}>\Delta t^*_{\mathcal{R}, l_{min}}$. The time distance of the JEPs occurrence are normalized by the reference time $t_{ref}=D/U$ ($\Delta t^*_{\mathcal{R}, l_{min}}=\Delta t_{\mathcal{R}, l_{min}}/ t_{ref}=l_{min}/St_a$ and ($\Delta t^*_{\mathcal{R}, l}/t_{ref}=l/St_a$). Superscripts $c-c$ and $c$ stand for the counter-clockwise and clockwise arm rotation.}
\def~{\hphantom{0}}
\begin{tabular}{lcccccccc}
\hline
$\mathcal{R}=m/n$ & $l_{min}$ & $\Delta \theta_{\mathcal{R},l_{min}}$  & $\Delta t^*_{\mathcal{R}, l_{min}}$ & $l_{int}$ $\left( \Delta \theta_{\mathcal{R},l_{int}}\right)$ & $l$ & $\Delta \theta_{\mathcal{R},l}$ & $\Delta t^*_{\mathcal{R}, l}$ & $St_l$\\[1pt]
\hline
$2/1$ & $1$ &$180^\textrm{o}$ & $2$& - & $1$ &$ 180^\textrm{o}$ & $2$ & $1/4$ \\
$5/2$ & $2$ &$72^\textrm{o}$ & $4$& - & $2$ &$ 72^\textrm{o}$ & $4$ & $1/20$ \\
$7/3$ & $2$ &$51.43^\textrm{o}$ & $4$& - & $2$ &$ 51.43^\textrm{o}$ & $4$ & $1/28$ \\
$12/5$ & $5$ &$30^\textrm{o}$ & $10$& - & $5$ &$ 30^\textrm{o}$ & $10$ & $1/120$\\
$25/11^{c-c}$, $25/14^{c}$ & $9$ &$14.4^\textrm{o}$ & $18$& $7 \left( 28.8^\textrm{o}\right)$ & $2$ &$ 43.2^\textrm{o}$ & $4$ & $3/100$ \\
$25/12^{c-c}$, $25/13^{c}$  & $2$ &$14.4^\textrm{o}$ & $4$& - & $2$ &$ 14.4^\textrm{o}$ & $4$ & $1/100$\\
$50/21^{c-c}$, $50/29^{c}$ & $19$ &$7.2^\textrm{o}$ & $38$ & $12 \left( 14.4^\textrm{o}\right)$, $7 \left( 21.6^\textrm{o}\right) $ & $5$ &$ 36^\textrm{o}$ & $10$ & $1/100$\\
$50/23^{c-c}$, $50/27^{c}$ & $13$ &$7.2^\textrm{o}$ & $26$ & $11 \left( 21.6^\textrm{o}\right)$ & $2$ &$ 28.8^\textrm{o}$ & $4$ & $1/50$\\
$250/101^{c}$ & $99$ &$1.44^\textrm{o}$ & $198$ & $52 \left(2.88^\textrm{o}\right)$, $47 \left(4.32^\textrm{o}\right) $ & $5$ &$ 7.2^\textrm{o}$ & $10$ & $1/500$ \\
$497/232^{c-c}$ & $15$ &$0.724^\textrm{o}$ & $30$ & -- & $2$ &$ 23.9^\textrm{o}$ & $4$ & $16/1000$ \\\hline
\end{tabular}
\end{table*}

Less freedom in the observation outcome exists for $\mathcal{R}$ with $m$, such that neither straight nor curved arms corresponding to $l_{min}$ can be identified. It may seem surprising that a small difference in $\mathcal{R}$ can lead to significantly different jet patterns. 
For instance, for $\mathcal{R}=25/11$ and $\mathcal{R}=25/12$, the jets are characterised by two spiral arms generated by JEPs spaced by $l_2\le l_{all}$, see Fig.~\ref{fig:helix_delta_n}. The important difference between these two solutions is the azimuthal distance between the vortices forming the spiral arms. The characteristic JEPs parameters ($\Delta\theta_{\mathcal{R},l}$, $\Delta t^*_{\mathcal{R}, l}=\Delta t_{\mathcal{R}, l}U/D$) for selected $\mathcal{R}$ values are given in Table~\ref{tab:JEP_param}. 
It shows that $\Delta\theta_{\mathcal{R},2}$ for  $\mathcal{R}=25/12$ is smaller than for $\mathcal{R}=25/11$, while the moments of occurrence of JEPs are the same ($\Delta t_{\mathcal{R}, l}=l/f_a$) in both cases. By analogy with the centres of mass (see Fig.~\ref{fig:vortices_path_single}), two successive vortices along a given arm can be represented by the points $P_1$ and $P_2$. Assuming that the vortices detach from the jet at $r_0=D/2$ (see Fig.~\ref{fig:vortices_path}) and flow along perfectly straight lines inclined at the angle $\alpha$, the distance between $P_1$ and $P_2$ at time $t$ can be estimated from the cosine theorem as  

\begin{equation}\label{eq:L_t}
    \mathcal{L}_{l}= |\mathbf{X}_{P_1}(t +\Delta t_{\mathcal{R}, l}) - \mathbf{X}_{P_2}(t)| =\sqrt{L_1^2 +L_2^2-2L_1L_2\cos(\Delta \theta_{\mathcal{R},l})}
\end{equation}

\noindent where ${L}_1=U_vt+{L}_0$, ${L}_2=U_v \left(t+\Delta t_{\mathcal{R}, l} \right)+{L}_0$ and ${L}_0=r_0/\sin(\alpha)$. For $l=m$, it follows from (\ref{eq:delta_theta}) that $\Delta\theta_{\mathcal{R},m}=0$, making the above formula time-independent and equivalent to $L_{\mathcal{A}-\mathcal{B}}$. For $l\ne m$, $\mathcal{L}_{l}$ differs from  $L_{\mathcal{A}-\mathcal{C}}$ and $L_{\mathcal{A}-\mathcal{D}}$.  Moreover, it increases with time and decreases with $\Delta\theta_{\mathcal{R},l}$. As will be discussed later, these dependencies influence the jet pattern seen by the observer. Regarding the solutions for $\mathcal{R}=25/11$ and $\mathcal{R}=25/12$, in the latter case $\Delta\theta_{\mathcal{R},2}$ is smaller (see Table~\ref{tab:JEP_param}) and thus the vortices are closer. It simplifies the identification of the arms but, on the other hand, intensifies the interactions between the vortices. In effect, they are more deformed. Additionally, the smaller $\Delta\theta_{\mathcal{R},2}$ translates to pronounced differences in the $\langle U_x\rangle$ distribution. For $\mathcal{R} = 25/11$, the 25 arms can be easily distinguished, whereas for $\mathcal{R} = 25/12$, the $\langle U_x \rangle$ isosurface exhibits a smooth conical shape. Similar differences occur between the jets characterized by five spiral arms obtained for $\mathcal{R}=12/5$ and $\mathcal{R}=250/101$. In the former case, $\Delta\theta_{\mathcal{R},5}$ is larger, which results in a larger distance between the vortices, more bent spiral arms and the $\langle U_x\rangle$ isosurface revealing 12 distinct arms. For $\mathcal{R}=250/101$ the $\langle U_x\rangle$ distribution closely resembles the one obtained for $\mathcal{R}=25/12$, though the instantaneous solutions differ significantly (5 spiral arms vs. 2 two spiral arms). 

Based on the discussion so far, it is clear that jets with spiral arms are a subset of multi-armed jets, and the appearance of spirals arises exclusively from a specific sequence of JEP occurrences. This sequence is also responsible for the apparent rotational motion of the arms. An example of this pseudo-rotation is presented in Fig.~\ref{fig:helix_023_027} (see supplementary movie 2) showing a temporal evolution of the solutions obtained for $\mathcal{R}=50/23$ and $\mathcal{R}=50/27$. Similarly to the cases with $\mathcal{R}=25/11$ and $\mathcal{R}=25/12$, the jets characterize the occurrence of two spiral arms, however, an intermediate value of $\Delta\theta_{\mathcal{R},l}$ (see Table~\ref{tab:JEP_param}) facilitates easier identification of non-overlapping vortices and their paths. From the presented figures, it is evident that the spirals appear 
rotating counter-clockwise ($\mathcal{R}=50/23$) and clockwise ($\mathcal{R}=50/27$), when looking at the jet from the top. The vortices look as if they were hung on a string set in a circular motion.
The direction of rotation of $l$-spiral arms depends on the position of successive $\theta_{\mathcal{R},k+l}$ and it can be shown that if $\mod_{2\pi}(l\frac{2\pi}{\mathcal{R}})<\pi$ the arms move clockwise.
Similar rules apply to multi-spiral jets. For instance, the jet with five spiral arms for $\mathcal{R}=250/101$ would also rotate clockwise, while the one for $\mathcal{R}=250/149$ would have the same shape but with the arms moving counter-clockwise.

The rotation frequency, defined as the frequency at which the spiral arms complete a full rotation cycle, is given by $f_{l} = \omega_{\mathcal{R},l}/2\pi = \Delta\theta_{\mathcal{R},l}f_a/2\pi l$. For the bifurcating jet ($\mathcal{R}=2$, $l=1$, $\Delta\theta_{\mathcal{R},l}=\pi$), this simplifies to $f_{l} = f_a/2 = f_r$. In terms of the Strouhal number, $f_{l}$ can be expressed as

\begin{equation}
    St_l=\frac{\Delta\theta_{\mathcal{R},l}}{2\pi}\frac{St_a}{l}=\frac{\Delta\theta_{\mathcal{R},l}}{2\pi}\frac{\mathcal{R}\,St_r}{l}
\end{equation}

\noindent From this, it follows that the arms rotate more slowly with the larger number of arms $l$ or smaller $\Delta\theta_{\mathcal{R},l}$ for a given $l$. 
In general, $St_l < St_a, St_r$, with its specific values provided in Table~\ref{tab:JEP_param}. For cases where $St_l \ll St_a$, the movement of the arms is very slow and may require prolonged observation to be noticeable. The non-dimensional time needed to complete one rotation cycle is $t_r=St_l^{-1}$.
The last row in Table~\ref{tab:JEP_param} refers to the solution obtained by~\cite{Li_et_al-JFM-2024}, which stimulated the present research. In those studies, the combined ML-BO procedure led to the solution with the axial and radial excitation frequencies corresponding to $St_a=0.497$ and $St_r=0.232$, which is equivalent to $\mathcal{R}=497/232$. With our current knowledge, we understand that detecting even one of the 497 arms is impossible, and the only observable jet pattern may consist of the spiral arms. Table~\ref{tab:JEP_param} indicates that only two cases are possible: a 15-armed or a 2-armed spiral jet, with only the latter being detectable and observable. Figure~\ref{fig:scheme}, briefly discussed in the Introduction section, shows these two arms. The period of the rotation of the arms, estimated by~\cite{Li_et_al-JFM-2024} based on instantaneous solutions and spectral analysis of the axial velocity signal, was equal to $t_r=62.5D/U$. Now, we know that this number relates to $St_l=0.016$ very precisely.

\begin{figure*}[htbp!]
\centering
\includegraphics[width=0.8\linewidth]{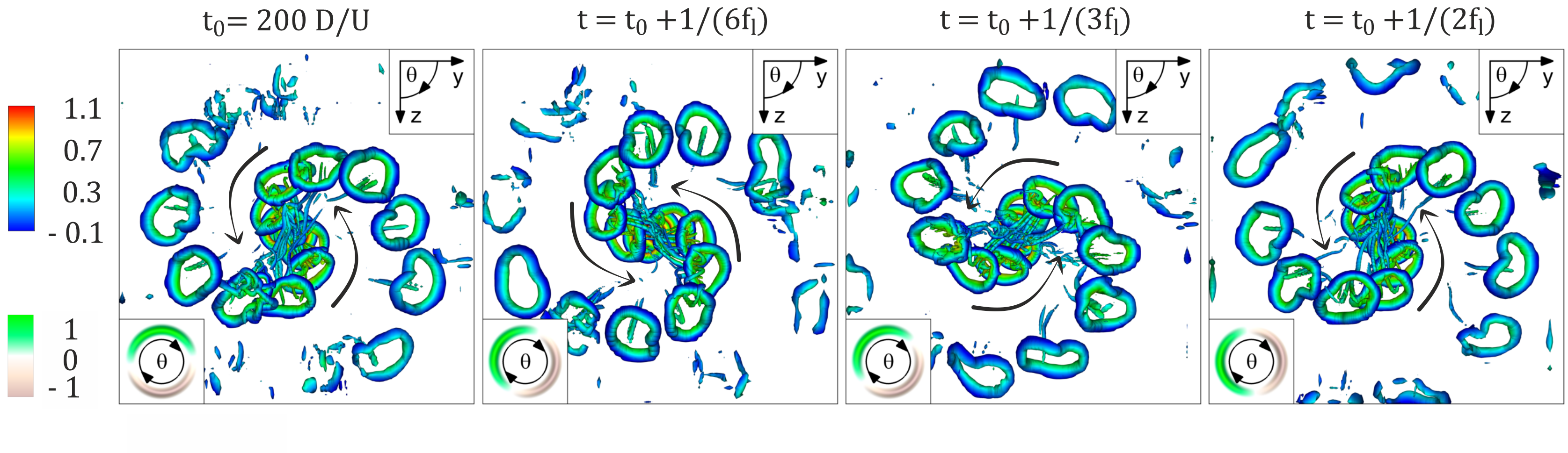}
\includegraphics[width=0.8\linewidth]{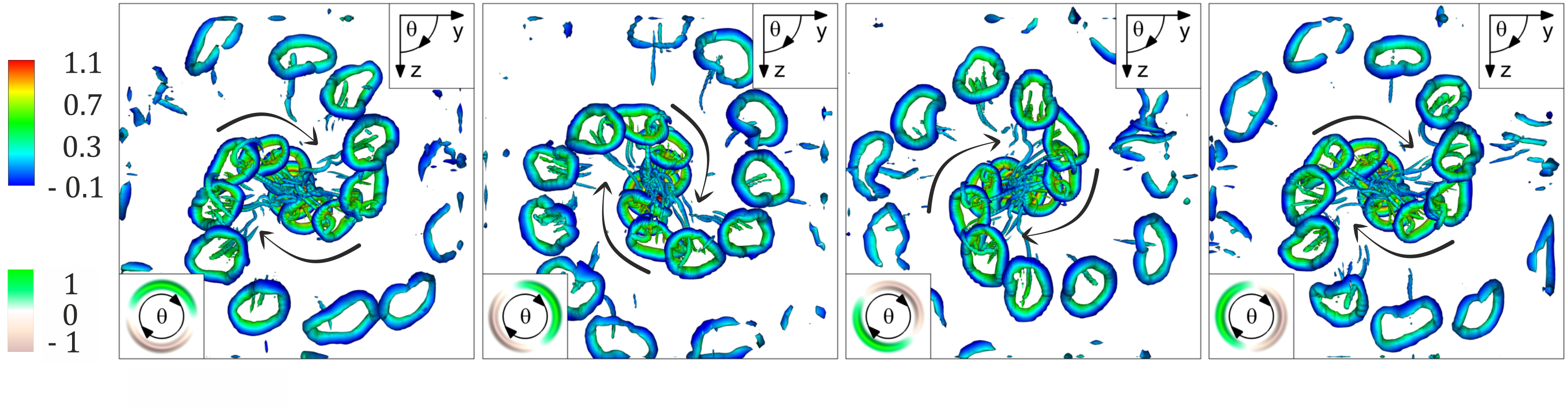}
\caption{Temporal evolution of the $Q$-parameter isosurface ($Q=1.0(U/D)^2$) coloured by the vertical velocity component normalised by the inlet jet velocity $(U_x/U)$ for cases with $\mathcal{R}=50/23$ (upper figures - counter-clockwise rotation marked by the arrow) and $\mathcal{R}=50/27$ (lower figures - clockwise rotation). Inset figures show $u_r/A_r$ contours at particular time instances.}
\label{fig:helix_023_027}
\end{figure*}

\begin{algorithm*}[ht!]
\caption{Simulation of the vortices motion}
\begin{minipage}{0.45\textwidth} 
\SetAlgoLined
\KwIn{$f_a, \mathcal{R}$: excitation parameters}
\KwIn{$L$: 2D domain dimension}
\KwIn{$t_{max}$: simulation time}
$\Delta t_{JEP} \gets 1 / f_a$\\
$t \gets 0$: current time\\
$k \gets 0$: vortex counter\\
$V[r,\theta] \gets \emptyset$: list of vortex coordinates\\
\While{$t \leq t_{max}$}{
    $k \gets k + 1$\\
    $V_k[r] \gets 1$\\
    $V_k[\theta] \gets \pi (4k - 3) / (2 \mathcal{R})$ (see Eq.~\ref{eq:theta_rk})\\
    $V \gets V \cup \{ V_k \}$\\
    \For{$i < k$}{
        $V_i[r] \gets V_i[r] + U_v \Delta t_{JEP}\sin(\alpha)$\\
        }
    Plot current positions of all 
    $t \gets t + \Delta t_{JEP}$\
}
\end{minipage}\hfill
\begin{minipage}{0.5\textwidth} 
    \includegraphics[width=0.8\linewidth]{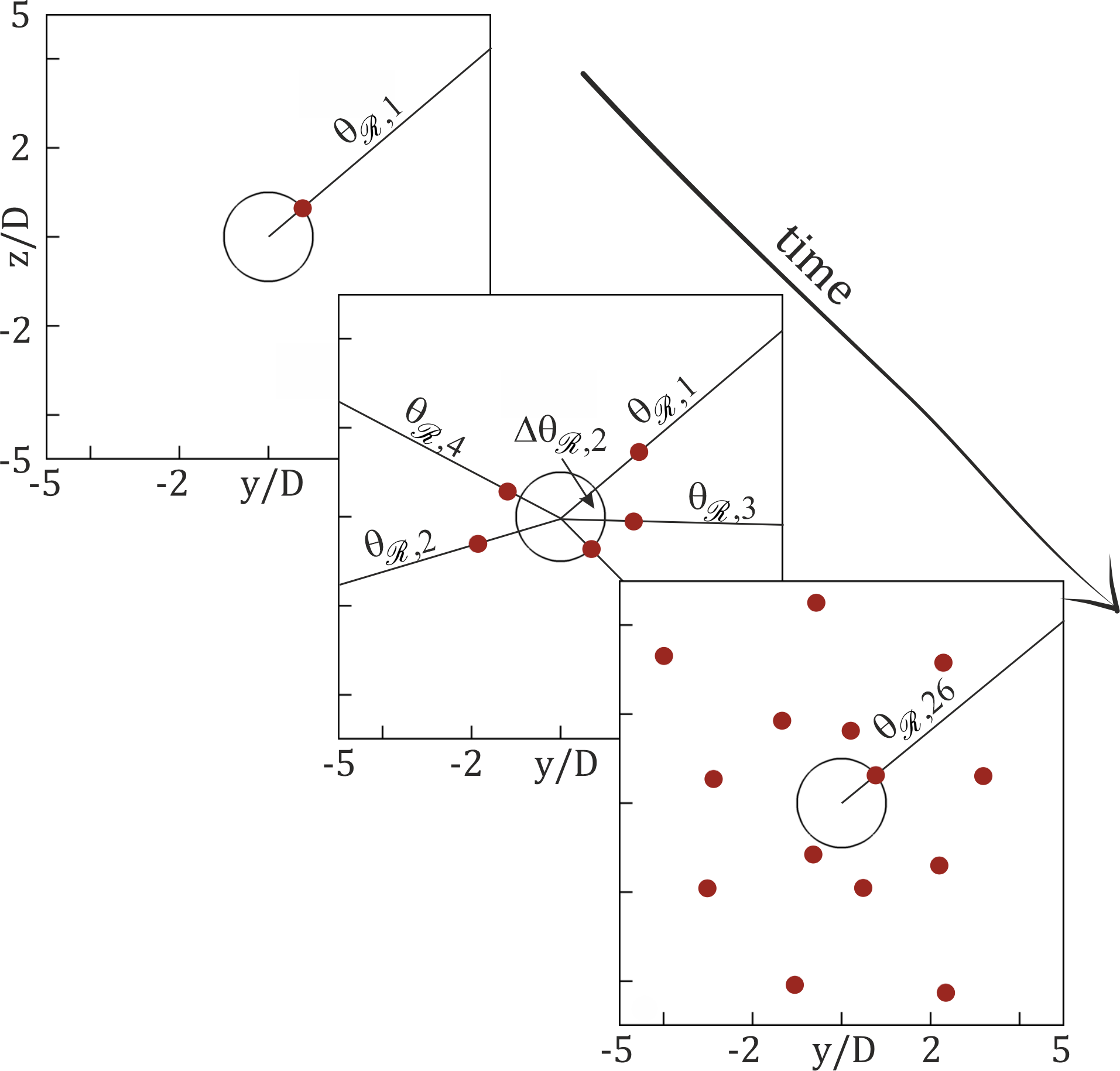} 
\end{minipage}
\end{algorithm*}

\subsection{Simple vortex motion model}
\begin{figure*}[t!]
\centering
\includegraphics[width=0.35\linewidth]{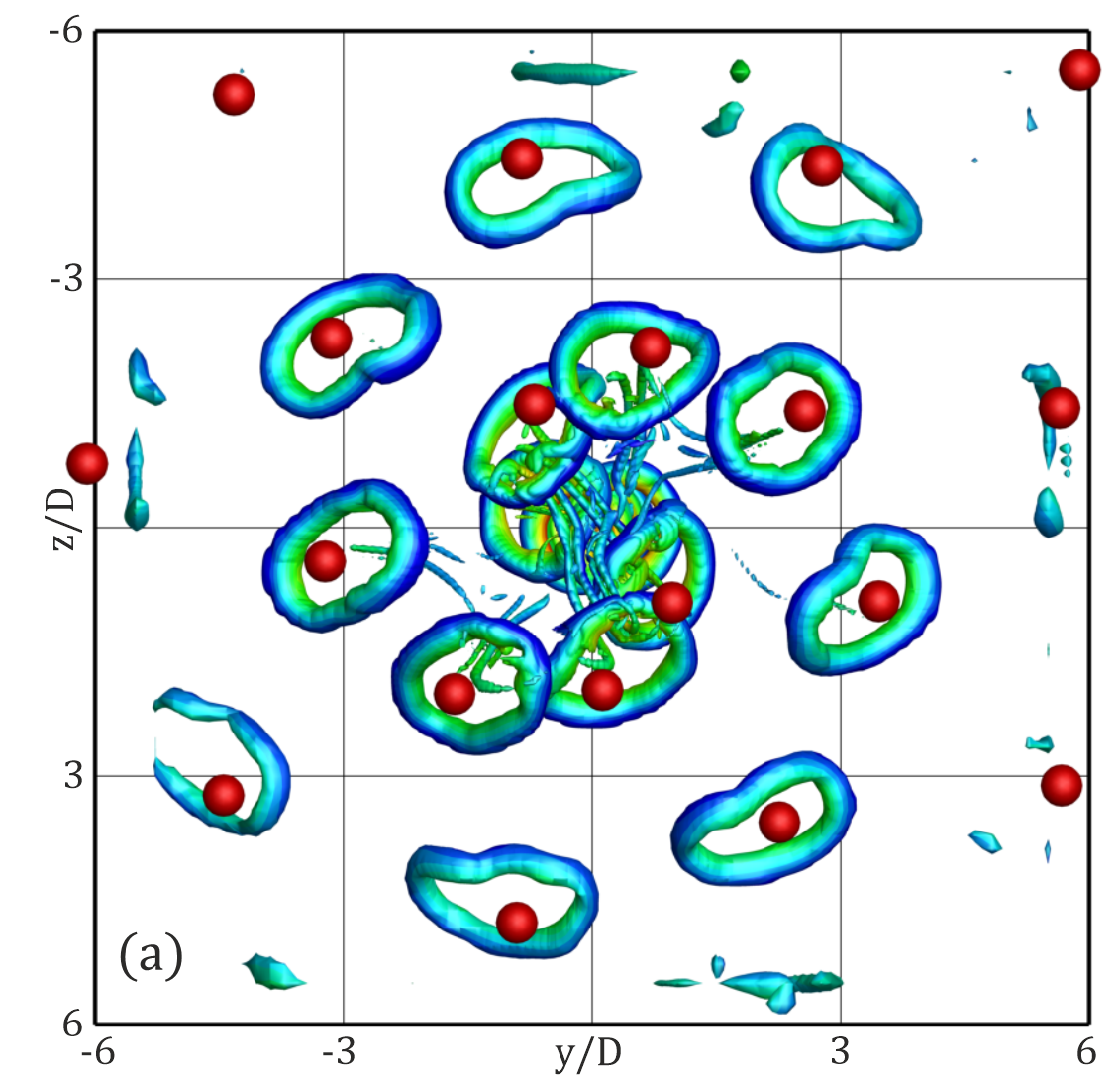}
\includegraphics[width=0.35\linewidth]{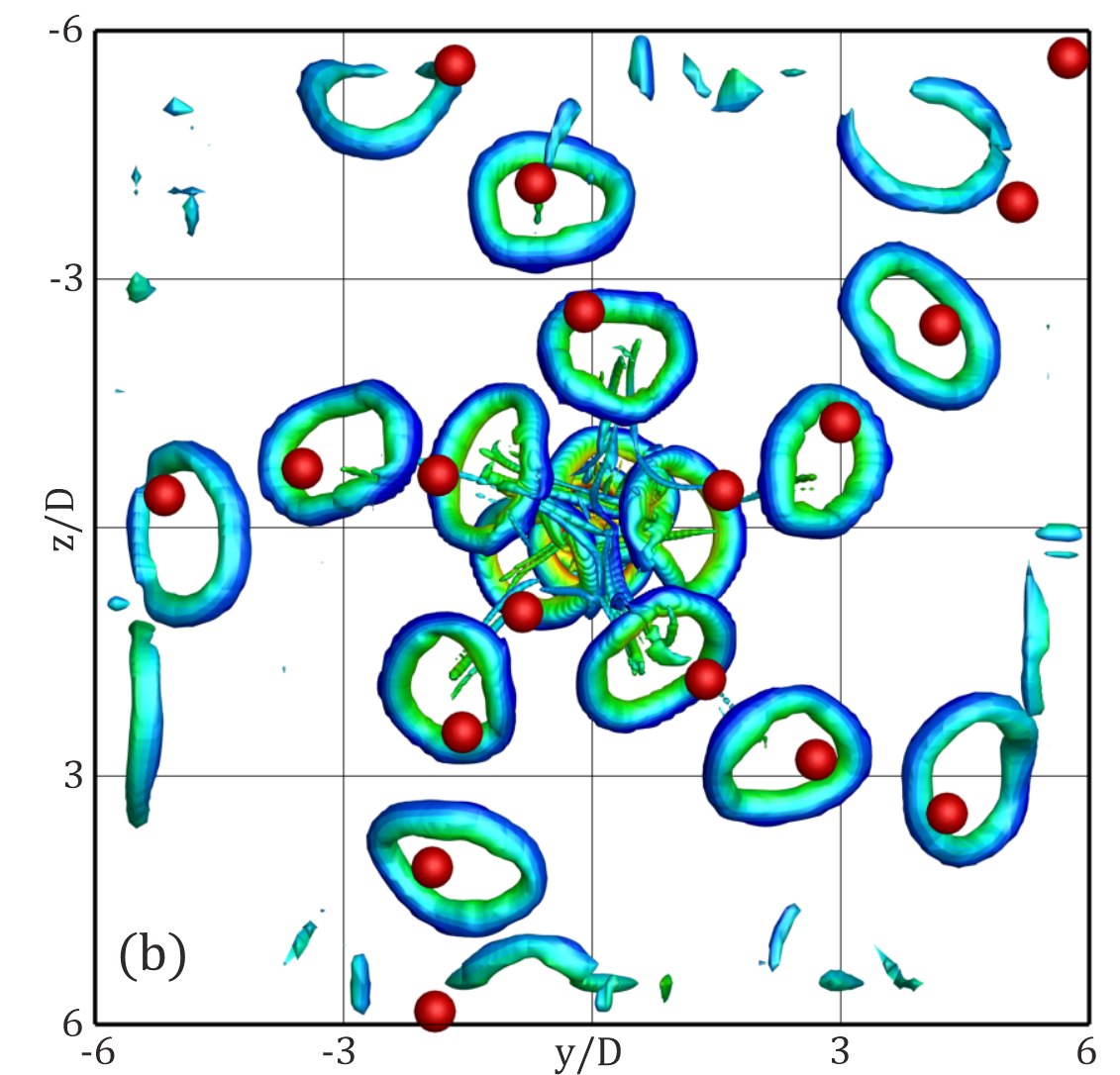}\\
\includegraphics[width=0.35\linewidth]{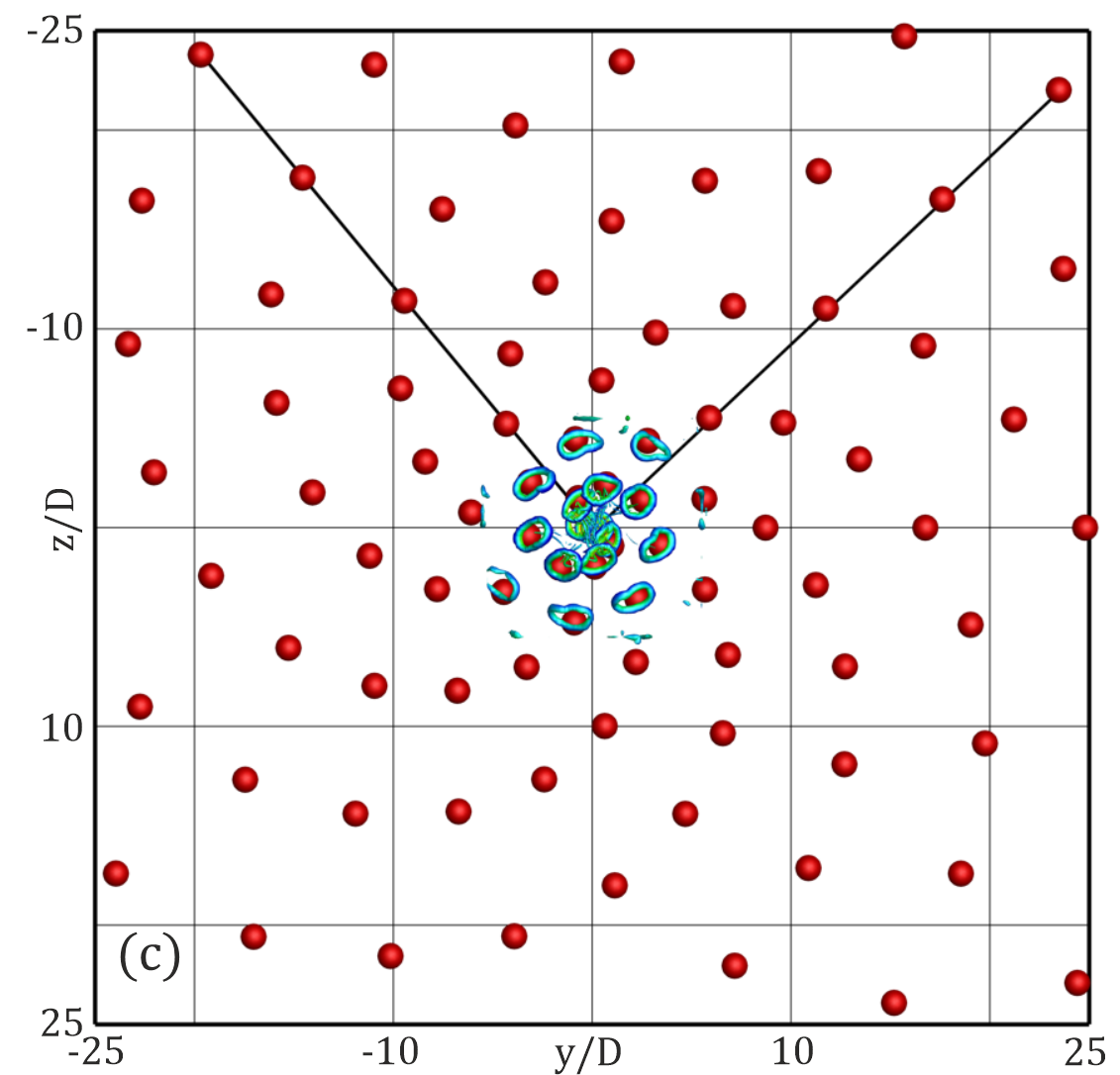}
\includegraphics[width=0.35\linewidth]{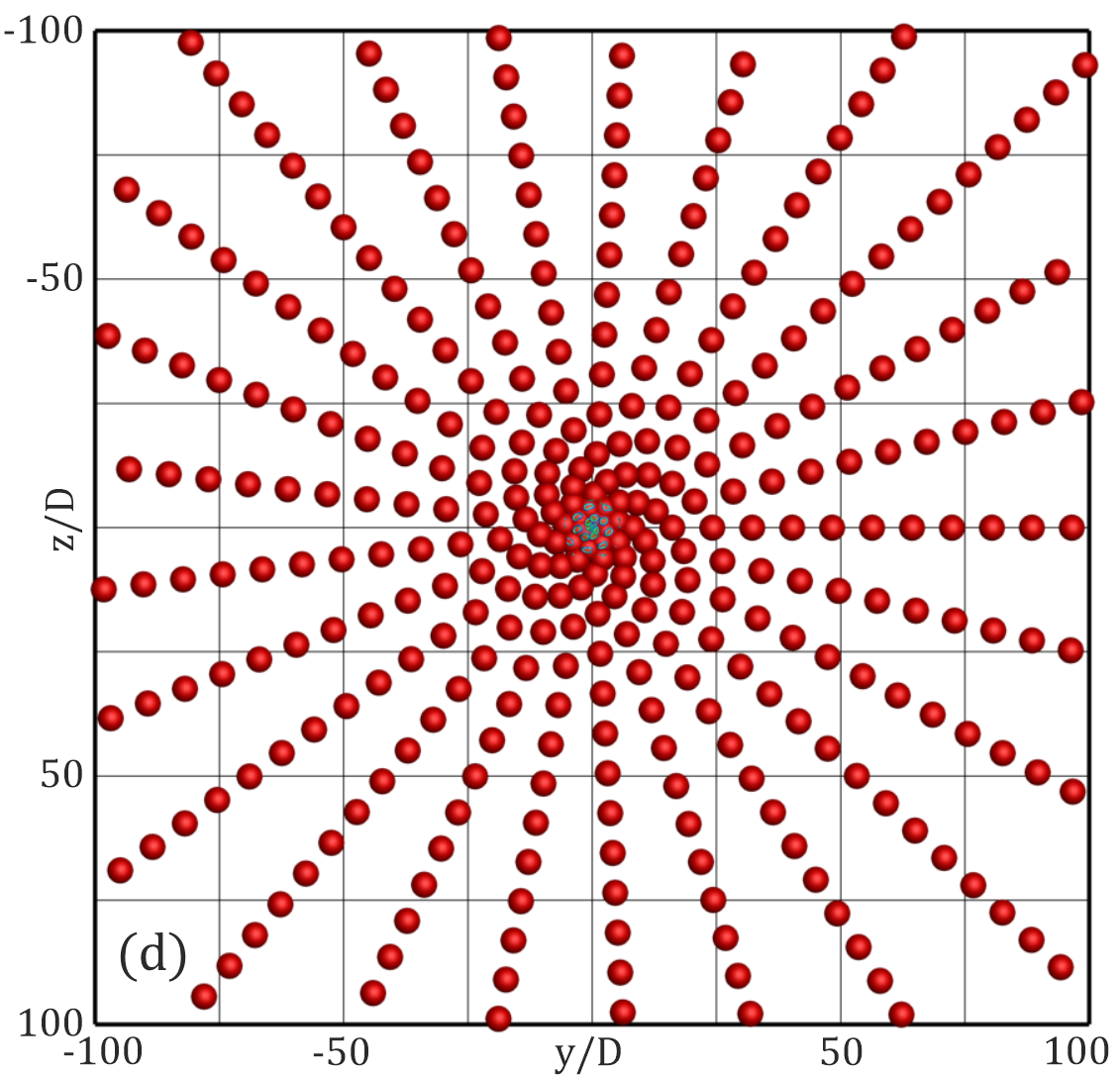}
\caption{Comparisons of the vortices distribution predicted by a simplified 2D model (red spheres) and LES for $\mathcal{R}=25/11$ (a) and $\mathcal{R}=250/101$ (b). Figures (c) and (d) show jet patterns observable in different sizes of the domain.}
\label{fig:model_LES}
\end{figure*}

The analysis presented above has shown that vortex structures detach from the main jet stream and move radially at specific $\theta_{\mathcal{R},k}$ defined by the JEPs locations. The assumption of $\mathcal{R}$ explicitly defines the number of straight arms; however, it does not allow for an immediate inference of what the observer would notice. By performing calculations that yield data like those in Table~\ref{tab:JEP_param}, one can identify likely jet patterns. Nevertheless, without conducting full 3D simulations, the exact shape of the existing arms would remain a matter of speculation. In this section, we formulate a simple model that reflects the vortices distribution seen above the jet, i.e., their locations are projected on the $y-z$ plane. To this end, we made a few simplifying assumptions: (1) the vortices have infinite lifetimes, (2) they are represented by virtual centres of mass and (3) the impact of turbulent flow is omitted. Additionally, based on what has been demonstrated so far, we assume that: (4) the vortices detach from the main jet stream at the radial distance $r=D$ and (5) the vortices move along the paths inclined to the main jet axis at the angle $\alpha=40^{\textrm{o}}$ (see Fig.~\ref{fig:vortices_path}) with the convection speed $U_v=U/4$. The procedure to generate vortices and simulate their motion is presented in Algorithm~1 along with a few snapshots showing an early phase of the arm formation for $\mathcal{R}=25/11$. Despite the exceptionally simple structure of the proposed model, it effectively predicts the vortical paths. Figures~\ref{fig:model_LES}a-b show the paths formed by the vortices position predicted by the simplified model on top of the results obtained by performing full 3D LES computations for the cases with $\mathcal{R}=25/11$ and $\mathcal{R}=250/101$ with two and five spiral arms. In both cases, the agreement is very good, particularly in the central part of the domain where the impact of the lateral boundaries on the 'real' vortices is minimal. The undisputed advantage of the simplified model is its ability to illustrate the jet's pattern, which could be observed if the observable domain were larger and the vortices did not dissipate. For instance, from Table~\ref{tab:JEP_param} we know that for $\mathcal{R}=25/11$ the maximum number of spiral arms equals $l=9$. From equation (\ref{eq:lspir}), one can calculate that, to observe nine spiral arms, each composed of at least three vortices, the distance $\mathcal{L}_{\mathcal{A}-\mathcal{C}}$ must be at least $13.5D$. If the domain were larger, the arms would consist of more vortices. This scenario is depicted in Fig.~\ref{fig:model_LES}c, where nine counter-clockwise arms are immediately identifiable. The diagonal lines indicate the directions of emerging straight arms. Probably, if they were not marked the observer would not noticed them. This is because in the observable $50D\times50D$ domain the distance between the vortices along the straight arms is larger than between the vortices along the spiral arms. According to equation (\ref{eq:L_t}), the latter increases with time, which means that at some moment $\mathcal{L}_{l=9}$ must become larger than $\mathcal{L}_{l=25}=\mathcal{L}_{\mathcal{A}-\mathcal{B}}$. This happens exactly at $t=171D/U$. In this time, the vortices move radially to the position $27.6D$. Hence, in domains larger than this, the dominant pattern of the jet should manifest as straight arms. Indeed, this is the case, as can be easily verified in Fig.~\ref{fig:model_LES}d. At a radius of approximately $30D$, the nine-spiral pattern disappears. Unfortunately, reproducing this analysis in reality is impossible due to the limited vortex lifetime.   









\section{Conclusions}

It has been demonstrated that the combined axial and radial excitation, as defined in equation~(\ref{eq:excitation}), with the frequency ratio $\mathcal{R}$ being a non-integer rational number, gives rise to a class of multi-armed jets characterized by vortical structures detaching from the main jet stream and moving radially. We formulated the rules enabling a priori assessment of whether, for a given $\mathcal{R}$, the arms of the jet can be noticed by an observer. The limiting factors for the simultaneous observation of all jet arms are the observation window's size and the vortical structures' lifetime. If the necessary conditions are not met (see equation (\ref{eq:marmed})), the time-evolving jet pattern resembles the rotating spiral arms with the vortices aligned along curved paths. For some values of $\mathcal{R}$, the number of spiral arms is univocally determined, e.g., five for $\mathcal{R}=12/5$ or two for $\mathcal{R}=25/12$. There are, however, cases for which the observed jet pattern may depend on an individual observer's perception.
For instance, with $\mathcal{R} = 25/11$, if the vortices' lifetime and size of the observation window are sufficiently large, nine, seven, and two spirals can be distinguished (see Table~\ref{tab:JEP_param}). The most noticeable pattern to an observer is likely the one where the vortices along a given set of spiral arms are closest to one another. 
It has been shown that the circular motion of the spiral arms results from the specific spatio-temporal distribution of the vortices. Although this motion is only apparent, its frequency can be precisely determined based on the assumed $\mathcal{R}$. Moreover, by a proper choice of $\mathcal{R}$ one can enforce clockwise or counter-clockwise rotation. 

A priori assessment of a standing-up jet pattern and its motion may not be straightforward, particularly for $\mathcal{R}$ values that result in multiple variants of spiral arm distributions. To simplify this process without requiring complex and time-consuming simulations, we developed a simple model that can be easily implemented using tools such as Python or MATLAB. It enables the visualization of the spatio-temporal distribution of vortices, which shows good qualitative agreement with the simulation results.

A natural question is \textit{What would happen to the jet if $\mathcal{R}$ were an irrational number?} In this case, 
as postulated by~\cite{ReynoldsParekh_AnnRev_2003}, one could expect that no vortex ring will exactly follow any other generated previously, resulting in their chaotic distribution. This, however, has never been verified and probably will not be verified in future. The reality is that irrational numbers cannot be represented with finite precision arithmetic in computer memory or in the settings of experimental devices. Consequently, if a jet splits into arms, the number of arms is finite and can be precisely determined by identifying the integers $m$ and $n$ that define $\mathcal{R}$. If they are impossible to see, the ordered spirals emerge. In other words, the only way to observe a chaotic motion of vortices is to add randomness to $\mathcal{R}$.

\bibliographystyle{ieeetr}
\bibliography{jfm}

\end{document}